\documentclass[aps,prb,twocolumn,superscriptaddress,floatfix,amsmath,
amssymb,aps]{revtex4}

\usepackage{graphicx}
\usepackage{dcolumn}
\usepackage{bm}
\usepackage{amsmath}
\usepackage{tabularx}
\usepackage{graphicx}
\usepackage{dcolumn}
\usepackage{bm}
\usepackage{amssymb}
\usepackage[export]{adjustbox}
\usepackage{xcolor}

\usepackage{xcolor}
\usepackage{soul}
\begin{document}

\title{Complex single-site magnetism and magnetotransport in single-crystalline Gd$_{2}$AlSi$_{3}$}

\author{Ram Kumar}
\email{ramphy21@umd.edu}
\author{Shanta R. Saha}\author{Jarryd Horn}\author{A. Ikeda}\author{Danila Sokratov}\author{Yash Anand}\author{Prathum Saraf}\author{Ryan Dorman}\author{E. Hemley}\affiliation{Maryland Quantum Materials Center, Department of Physics, University of Maryland, College Park, MD 20742, USA}
\author{K. K. Iyer}
\affiliation{Tata Institute of Fundamental Research, Homi Bhabha Road, Colaba, Mumbai 400005, India}
\author{Johnpierre Paglione}
\email{paglione@umd.edu}
\affiliation{Maryland Quantum Materials Center, Department of Physics, University of Maryland, College Park, MD 20742, USA}
\affiliation{Canadian Institute for Advanced Research, Toronto, Ontario M5G 1Z8, Canada}

\date{\today}

\begin{abstract}
We present a detailed investigation of single-crystal samples of the magnetic compound Gd$_{2}$AlSi$_{3}$, which crystallizes in the $\alpha$-ThSi$_2$ type tetragonal structure. We report the temperature and magnetic field dependence of the magnetic susceptibility, magnetization, heat capacity, electrical resistivity, and magnetoresistance for magnetic fields applied along both the tetragonal $c$-axis and in the basal $ab$-plane. X-ray diffraction measurements confirm a centrosymmetric, $I4_{1}/amd$ space group of the crystal structure. Despite single-site occupancy of the Gd position in this tetragonal structure, we identify two successive antiferromagnetic phase transitions at Ne\'el temperatures 32~K and 23~K via magnetic susceptibility, heat capacity and transport measurements, as well as a complex magnetic interaction with a magnetic anisotropy that plays an important role in the direction-dependent transport response. Our identification of multiple magnetic phases in Gd$_{2}$AlSi$_{3}$, where Gd is the only magnetic species, helps to elucidate the field-induced skyrmionic behavior in the Gd-based intermetallic compounds. 
\end{abstract}

\maketitle

\section{Introduction}
 
In recent decades, the Ruderman-Kittel-Kasuya-Yosida (RKKY) interaction, serving as an indirect exchange mechanism in metallic magnetic materials, has gained recognition for its role in facilitating magnetic interactions among various magnetic sites. It has been recognized that the magnetic interaction mediated by itinerant electrons can result in unconventional and conceptually intriguing consequences, such as distinctive magnetic ground states \cite{Batista_2016, PhysRevApplied.11.044046} and abnormalities in electrical transport \cite{PhysRevLett.117.206601, PhysRevB.101.184432}, especially within metallic environments. Geometrically frustrated magnetism, arising from the spatial configuration of magnetic ions, is a contemporary area of investigation in condensed matter physics. This form of frustration is recognized for its capacity to give rise to a diverse array of intriguing magnetic states \cite{doi:10.1143/JPSJ.42.76, PhysRevB.12.5007, PhysRevB.65.180401, Shaginyan, Levy_2008}. It is evident that one observes indications of the competition between ferromagnetic (FM) and antiferromagnetic (AFM) interactions in the measured properties. The magnetic ground state resulting from the RKKY interaction can experience such frustration, a condition that has been suggested to give rise to magnetic skyrmions \cite{PhysRevB.97.054408} — an exotic form of magnetic textures that currently garners significant interest.

A good number of rare-earth-based ternary intermetallic compounds, R$_{2}$TX$_{3}$ (R = rare earth, T = transition metal, and X = Si, Ge), have attracted significant attention both from fundamental and application points of view due to the manifestation of various interesting physical properties associated with complex structure-property relationships \cite{gordon1997substitution,10.1103/physrevb.60.12162, tien20002,majumdar2001magnetic, szlawska2010experimental, tang2011crystallographic,2015PhRvB..91c5102B, pakhira2016large, kurumaji2019skyrmion, PhysRevB.101.144440, szytula1999magnetic, li2001magnetic, smidman2019magnetic, litzbarski2020synthesis, pan2013structures}. Over the past several decades, a lot of research activities have been conducted to understand various phenomena associated with the delocalization of the 4$f$ electrons (for example Ce, Eu, and Yb) which led to path-breaking discoveries in condensed matter physics. On the other hand, compounds with heavy rare-earth members (e.g.,  Gd, Tb, and Dy) with localized 4$f$ electrons have attracted less attention until recently. In this respect, a standout example is found in Gd$_{2}$PdSi$_{3}$, which crystallizes in the AlB$_2$-derived hexagonal crystal structure and is known to exhibit anomalous transport behaviors such as Kondo-like features in electrical resistivity \cite{R.Mallik_1998}. Notably, a Hall anomaly, characteristic of a non-trivial topology in electronic structure, was reported two decades ago, preceding its recognition in metals \cite{sampathkumaran2019report, 10.1103/physrevb.60.12162}.

Gd$_{2}$PdSi$_{3}$ represents the first centrosymmetric system displaying magnetic skyrmion behavior across two metamagnetic transitions in the magnetically ordered state\cite{kurumaji2019skyrmion, PhysRevLett.129.137202}.~A more recent report highlighted the discovery of polycrystalline Gd$_{2}$AgSi$_{3}$, which crystallizes in the $\alpha$-ThSi$_2$ tetragonal structure, but still exhibits magnetic and transport properties similar to those of hexagonal Gd$_{2}$PdSi$_{3}$\cite{SAHU2021107214}.~Here, we report the first synthesis and characterization of a new ternary compound, Gd$_{2}$AlSi$_{3}$, which replaces the transition metal position with aluminum to probe the role of the transition metal in the exotic properties in these systems.  We present detailed temperature and field-dependent magnetism and transport properties, providing the first such study in the R$_{2}$AlSi$_{3}$ family.

\section{EXPERIMENTAL DETAILS}

Single crystals of Gd$_{2}$AlSi$_{3}$ were grown using the high-temperature self-flux technique with molten Al as a solvent. Gd metal with 99.9$\%$ purity, Al (shot) and Si (lump) with purities greater than 99.999$\%$, were obtained from Alfa-Aesar.The starting materials were taken in a molar ratio of Gd:Al:Si = 2:85:3. Typically, the reactions were carried out in 2-cm$^3$ alumina crucibles, which were encapsulated in evacuated fused silica jackets by flame sealing. The temperature profile: ramping to 1100$^{\circ}$C with the rate 100$^{\circ}$C/hr, homogenization for up to 24 hr, cooling to 850$^{\circ}$C (above the melting point of Al, 660$^{\circ}$C) at 2$^{\circ}$C/hr rate. At 850$^{\circ}$C, the molten Al was removed by centrifugation. The structural analysis was done with an x-ray diffractometer (Rigaku, MiniFlex600) where X-ray diffraction (XRD) patterns were taken on a powder with monochromated Cu-K$\alpha$ radiation ($\lambda \sim 1.5406~$\AA. The powder sample for XRD was prepared from a single crystal, which was subsequently crushed and ground in agate mortars. The collected XRD data were used for phase identification by the Rietveld refinement\cite{Rietveld_a07067} technique using FullProf software packages. The homogeneity and chemical composition were verified using a scanning electron microscope (SEM) with an energy-dispersive x-ray spectroscopy (EDS) analyzer, provided by JEOL.

Temperature and field-dependent magnetic measurements were carried out using a Quantum Design SQUID-VSM and Dynacool in the temperature and field ranges of 1.8–300~K and 0-14~T, respectively. The $\chi$(T) measurements were conducted in the zero-field-cooled (ZFC) mode, involving the cooling of the sample to the base temperature without applying a magnetic field. The magnetization M(T) was then measured during the warming cycle after the application of a field. For every magnetic measurement, the magnetic field was oscillated to zero before each measurement to minimize residual magnetic flux trapped in the superconducting solenoid. A Physical Properties Measurement System (PPMS, Quantum Design) was used to measure the specific heat C$_P$(T) and electrical resistivity measurements $\rho$(T) in the temperature and field ranges 1.8–300 K and 0-14 T, respectively. $\rho$(T) measurements were performed on as-grown single crystals. In all resistivity measurements, the magnetic field was oriented transverse to the current direction.

\section{Results and Discussion}

\begin{figure}[h]
	\centering   \adjincludegraphics[height=12cm,trim={5cm 0 7cm 0},clip]{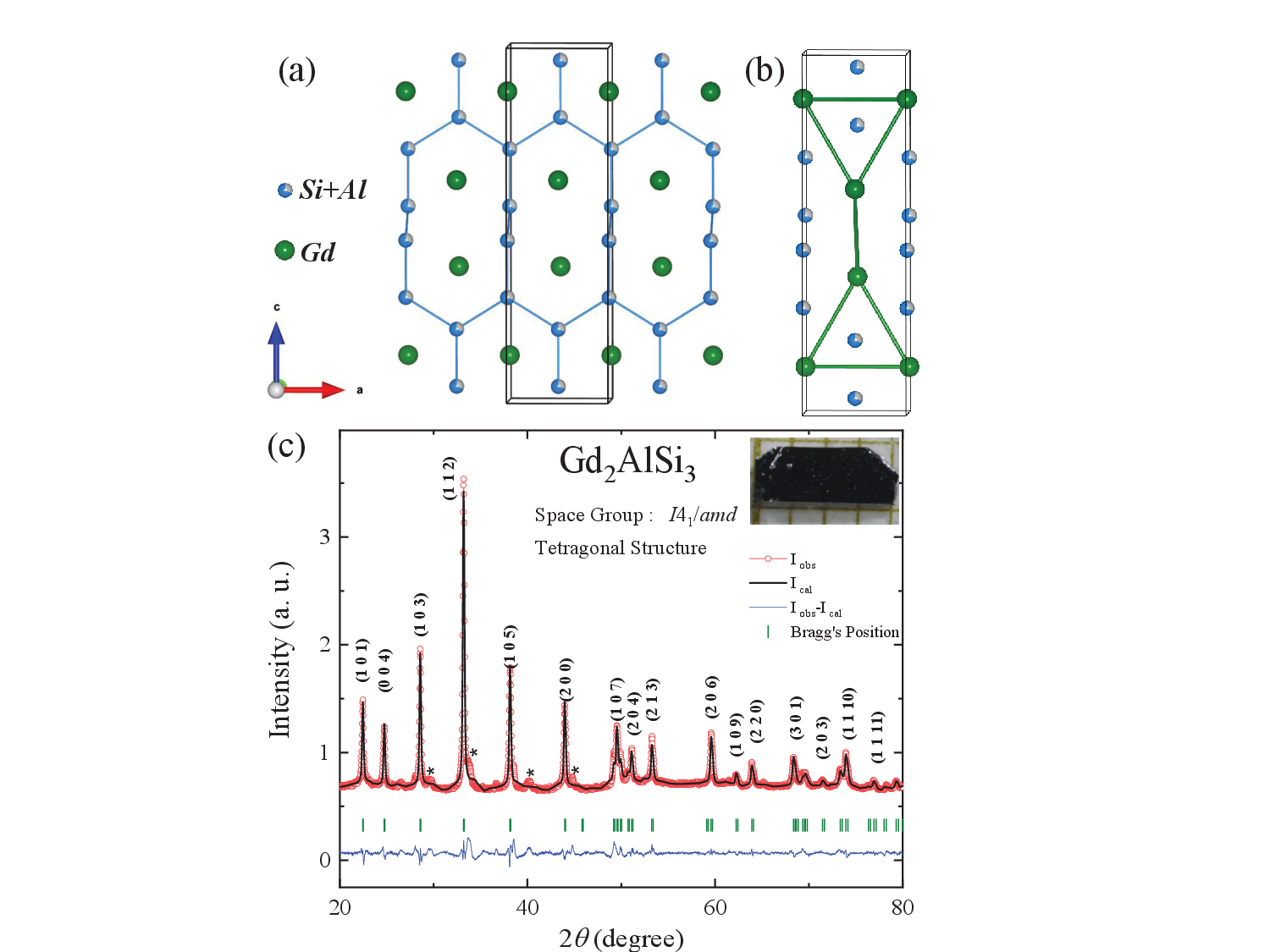}
	\caption{(a) A tetragonal crystal structure of Gd$_{2}$AlSi$_{3}$, (b) The triangular arrangements of Gd-atoms along the c-direction. (c) Rietveld refinement of powder x-ray diffraction data for Gd$_{2}$AlSi$_{3}$ with the experimental data (Red symbols) and the black line representing the calculated data along with asterisks for impurities. A set of vertical green bars represents the Bragg peak positions of the tetragonal ${\alpha}-$ThSi$_2$ type structure. Inset shows the as grown Gd$_{2}$AlSi$_{3}$ crystal.}
\end{figure}

\begin{figure}[h]
	\centering   \adjincludegraphics[height=13cm,trim={0.4cm 0.6cm 0.25cm 0},clip]{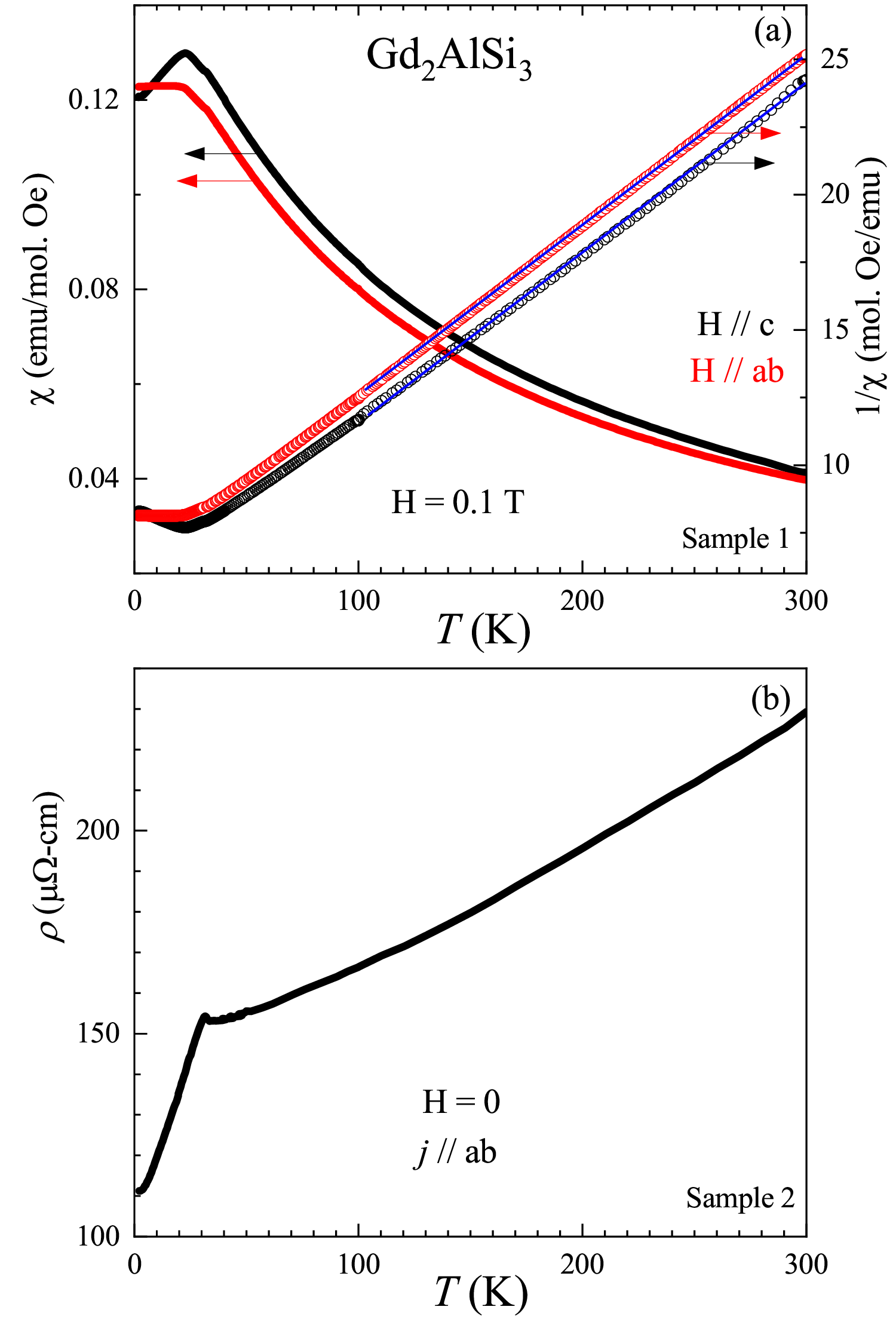}
\caption{(a) Magnetic susceptibility, $\chi$(T), along with inverse susceptibilities of Curie-Weiss fits for both directions with $\mu_{eff}$ = 8.06 $\mu_{B}$/Gd and $\theta_P$ = -91 K for $H \parallel c$ and $\mu_{eff}$ = 7.96 $\mu_{B}$/Gd and $\theta_P$ = -98.4 K  for $H \parallel ab$. (b) Temperature dependence of the zero fields in-plane resistivity, $\rho$ of Gd$_{2}$AlSi$_{3}$.}.
\end{figure}

\begin{figure}[h]
\adjincludegraphics[height=13cm,trim={0.4cm 0cm 0.15cm 0cm},clip]{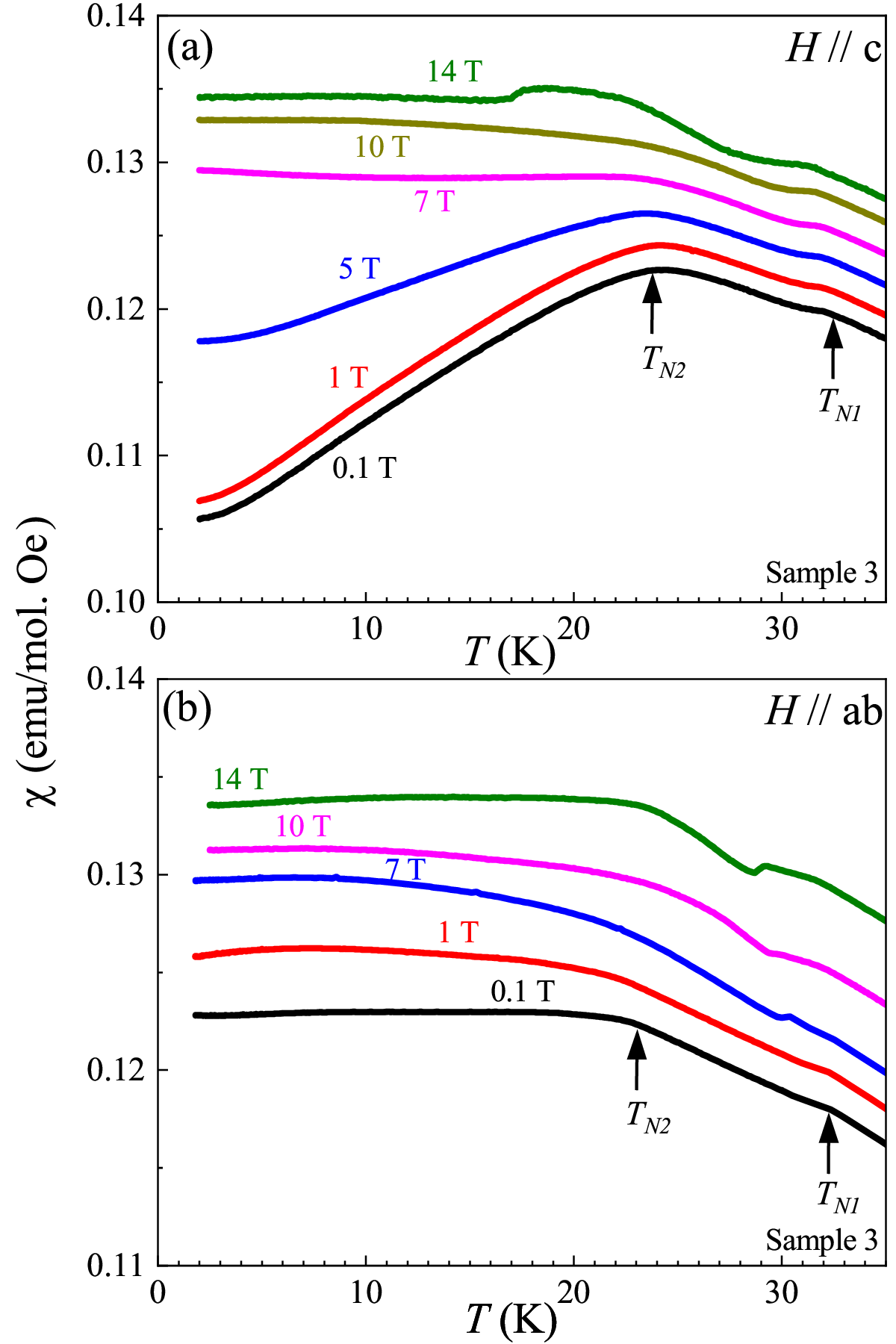}
    \caption{
    Temperature dependence of dc-magnetic susceptibility of Gd$_{2}$AlSi$_{3}$ measured under different magnetic fields between 0.1 T to 14 T for both $H \parallel c$ (a) and for $H \parallel ab$ (b) directions. Two black vertical arrows show two distinct magnetic transition temperatures, $T_{N1}$ and $T_{N2}$.
    }
\end{figure}

\subsection{Structural properties }

Figs.~1(a,b) present the chemical crystal structure and Fig.~1(c) presents the Rietveld refinement of the powder XRD data obtained for Gd$_{2}$AlSi$_{3}$ using the $I4_{1}/amd$ space group, which is a tetragonal structure of $\alpha$-ThSi$_2$ type having only single Gd-sites (different from Ce$_{2}$PdGe$_{3}$ with $\alpha$-ThSi$_2$ structure, which has two distinct Ce sites \cite{2015PhRvB..91c5102B}). The lattice parameters are found to be 4.116(1) Å for the $a$-axis and 14.386(1) Å for the $c$-axis. The asterisks show the possible impurity phase due to SiO$_{2}$ and Al$_{2}$O$_{3}$. From the structural analysis, we found that the smallest Gd–Gd bond length in Gd$_{2}$AlSi$_{3}$ is 4.1157(1) Å, which is larger than the expected Gd–Gd bond (atomic radii) of ${\sim}$3.6 Å. This observation suggests a rather weak interaction between the Gd atoms in this system. However, the triangular arrangement of Gd atoms (as shown in Fig. 1b) confirms the possibility of geometrically frustrated magnetism in the system. EDS analysis performed on cleaned surfaces of a single-crystal sample confirms the stoichiometric ratio, with the atomic composition of 34.14(2)\% Gd, 15.99(1)\% Al, and 49.87(1)\% Si. The compositions derived from EDS data align well with the refinement results from XRD data.

\subsection{Magnetic Susceptibility} 

Fig.~2 shows the combined plot of magnetic susceptibility and electrical resistivity ($\rho$) as a function of temperature. Temperature-dependent dc-magnetic susceptibilities $\chi(T)$ for both directions along with inverse magnetic susceptibilities, measured under ZFC mode with an applied field of $H$ = 0.1 T is shown in Fig.~2(a). It is observed that $\chi(T)$ exhibits two AFM-like anomalies below 50 K at Ne\`el temperatures $T_{N1}$= 32 K and $T_{N2}$~=~23~K. Curie-Weiss fits of inverse susceptibilities for both directions are shown on the right axis of Fig.~2(a). These fits lead to values of the effective magnetic moments ($\mu_{eff}$) and Weiss temperatures ($\theta_P$) of 8.06 $\mu_{B}$/Gd and  = -91 K for $H \parallel c$, and 7.96 $\mu_{B}$/Gd and -98.4 K for $H \parallel ab$, respectively. The obtained $\mu_{eff}$ values are comparable to the theoretical free-ion value for Gd$^{3+}$, which is 7.94 $\mu_{B}$. The negative sign of $\theta_P$ indicates that the dominant exchange interactions are of AFM type in the system. Since the magnitude of $\theta_P$ is higher than the observed ordering temperature in Gd$_{2}$AlSi$_{3}$, we infer strong competition with FM correlations. The temperature dependence of electrical resistivity of  Gd$_{2}$AlSi$_{3}$ under zero applied magnetic field is shown in Fig.~2(b), clearly demonstrating metallic behavior along with sharp anomalies at the higher ordering temperature $T_{N1}$ = 32 K.

To get insight into these two magnetic phase transitions, detailed temperature-dependent magnetization was measured at different applied magnetic fields for both $H \parallel c$ and $H \parallel ab$, as shown in Fig.~3(a) and (b), respectively. In comparison to the low-field data, the higher-field magnetization data exhibits distinct magnetic transitions as seen in Fig 3(a) and (b). Another interesting finding is the magnetic susceptibility behavior in the region $T_{N2}<T<T_{N1}$ where $T_{N2}$ shifts towards lower temperatures with an increasing applied magnetic field. At fields above 10~T, there is a marked change in the behavior of $\chi(T)$ going through $T_{N2}$, which is not the case for $T_{N1}$ (see Fig.~3(a)), as exemplified by the enhancement of $\chi(T)$ in 14~T in this range. These phenomena may be due to competing AFM-FM exchange interactions in the system that become prevalent in higher fields.

\begin{figure}[h]
\adjincludegraphics[height=15cm,trim={0.05cm 0.1cm 0cm 0.2cm},clip]{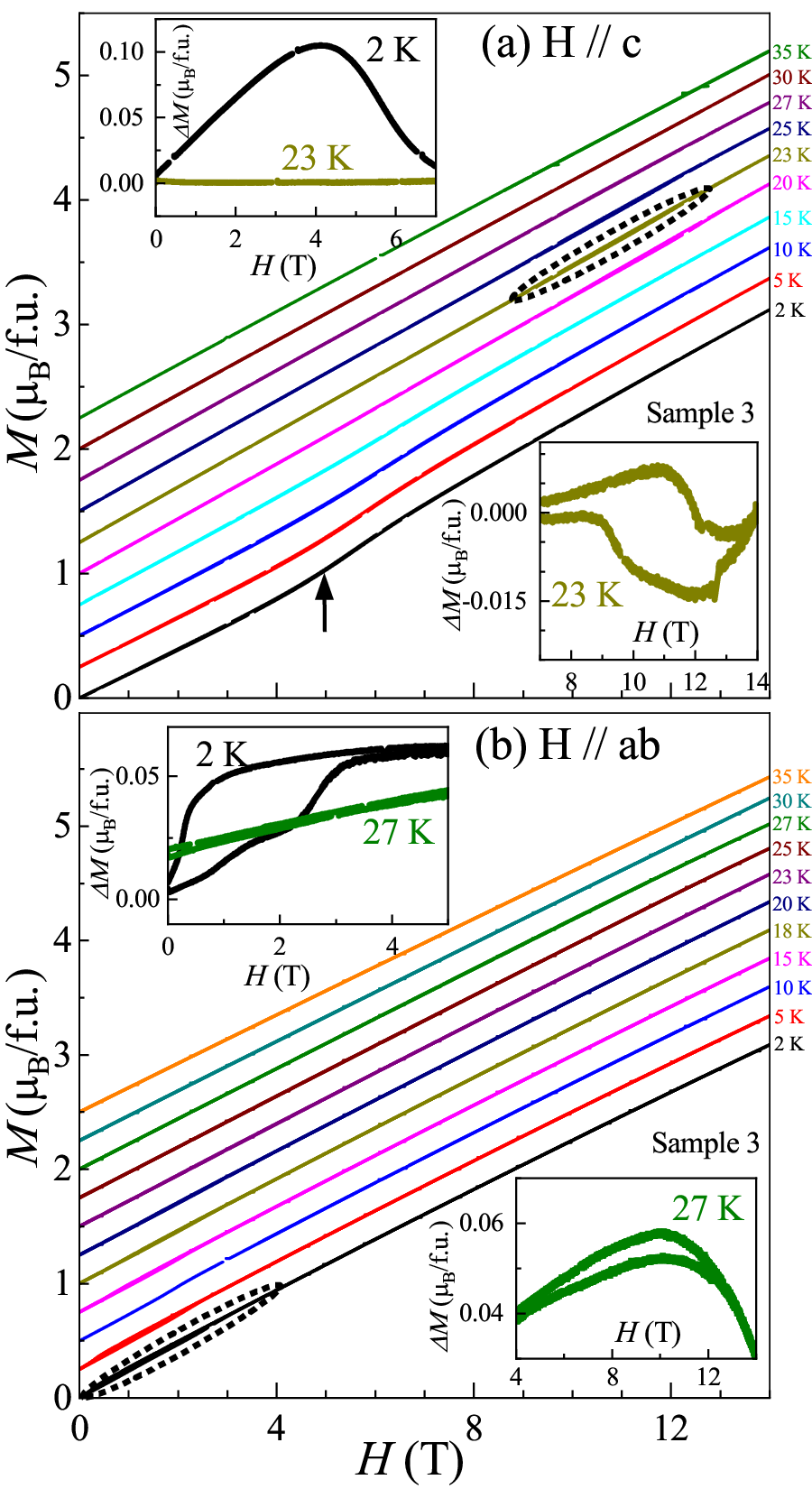}
    \caption{Magnetic field dependence of magnetization (M) of Gd$_{2}$AlSi$_{3}$ measured at different temperatures for $H \parallel c$ and $H \parallel ab$ as shown in (a) and (b), respectively. These plots are constantly shifted upwards for clarity of features like a field-induced-magnetic transition indicated by a vertical arrow for 2 K and the presence of small hysteresis indicated by dotted circles in both panels. All the insets show $\Delta$M versus H plots to catch such exotic features which are discussed in the text.}
\end{figure}

\begin{figure}[h]
	\adjincludegraphics[height=11cm,trim={0.18cm 1.3cm 1.2cm 2.5cm},clip]{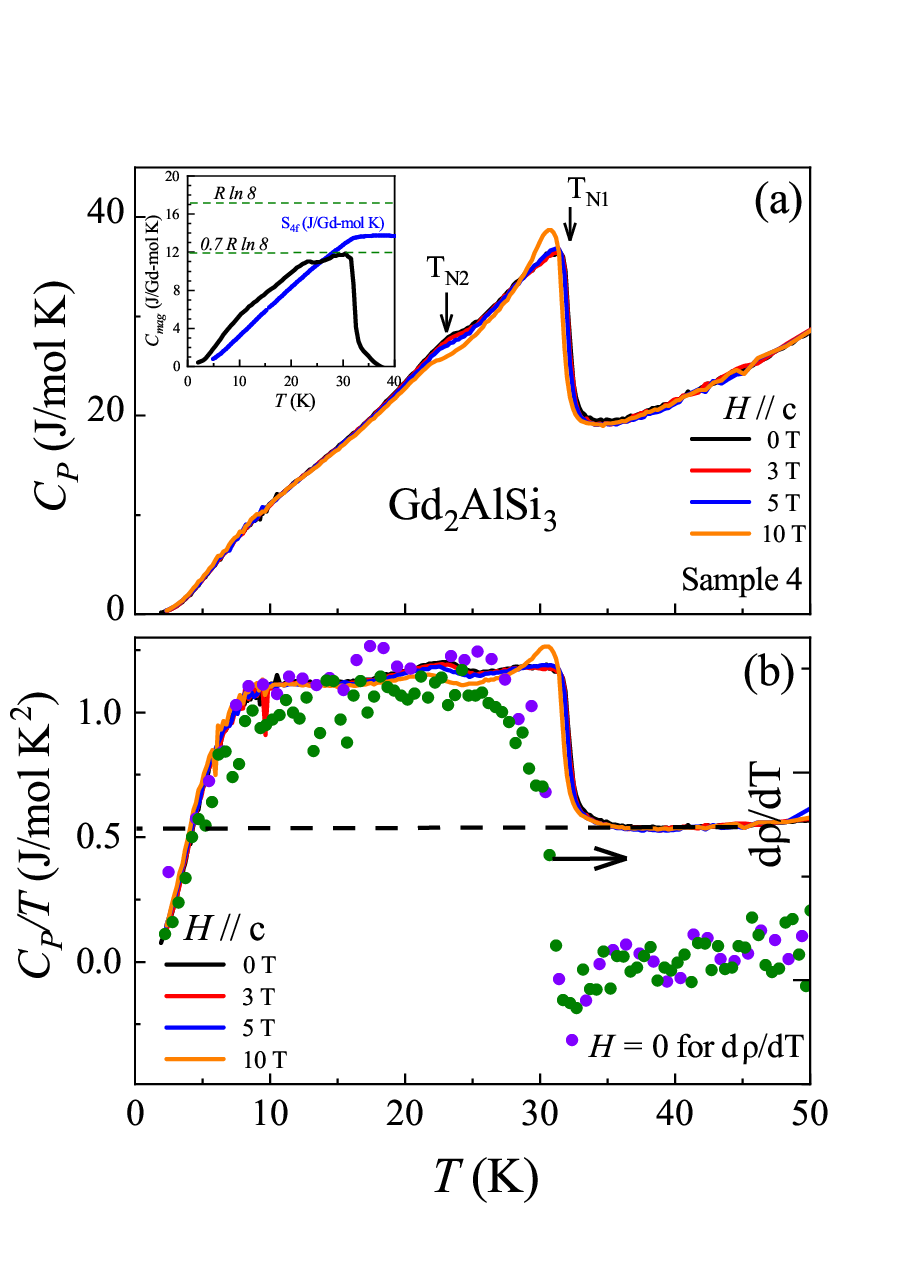}
    \caption{(a) Temperature dependence of Heat Capacity of Gd$_{2}$AlSi$_{3}$ measured under different applied magnetic fields up to 10 T. The inset shows the magnetic (4\textit{f} contribution) entropy for the system. (b) The C$_{P}(T)$/T vs T for a better view of these magnetic phase transitions along with d$\rho$/d\textit{T} (on right-axis) for both directions. A dotted line is drawn through the linear region to calculate the \textit{${\gamma}$} value.}
 \end{figure}

\begin{figure*}[ht]
    \centering
	\adjincludegraphics[height=7cm,trim={0.4cm 0.9cm 0.15cm 0.1cm},clip]{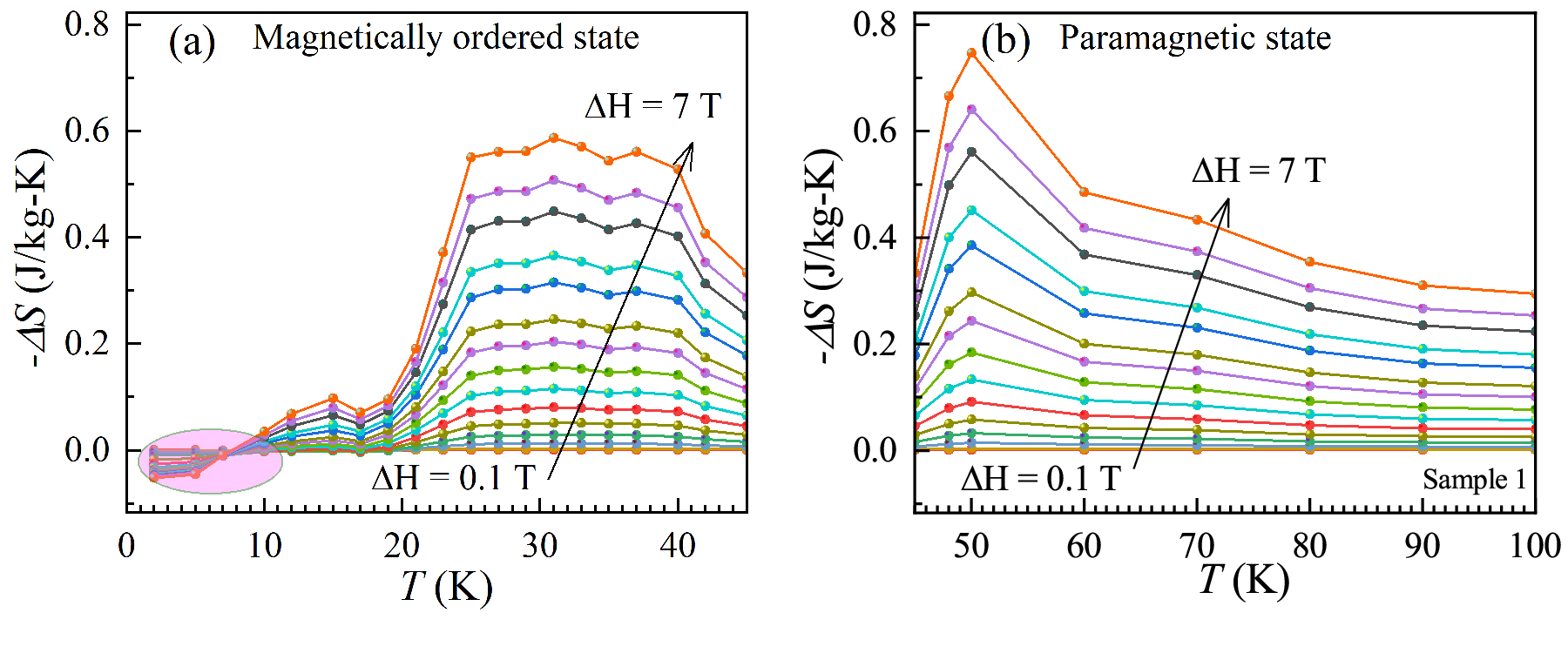}
    \caption{ Isothermal entropy change as a function of temperature for various final fields with initial zero external field obtained from isothermal magnetization data using Maxwell's relations, (a) around magnetic ordering and, (b) in the paramagnetic state.}
 \end{figure*}

 \begin{figure*}[ht]
    \centering
	\adjincludegraphics[height=7.1cm,trim={2.5cm 0.5cm 2cm 0.5cm},clip]{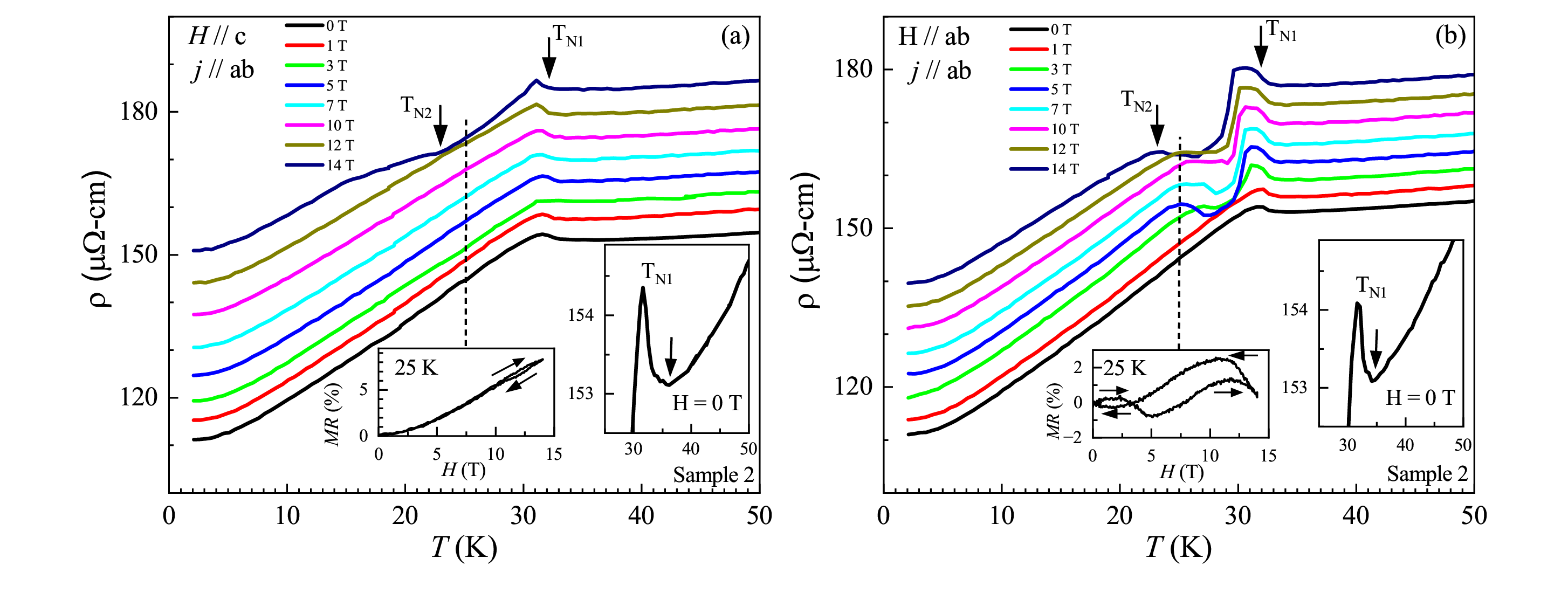} 
    \caption{Electrical resistivity as a function of temperature
(1.8–50 K) for Gd$_{2}$AlSi$_{3}$ in the presence of different external magnetic fields in both $H \parallel c$ (a) and $H \parallel ab$ (b) while the current flows in the same direction along $ab$-plane. We have constantly shifted the curves upwards for clarity. The respective insets along with dotted lines show the MR at that particular temperature in the respective plots while the other insets indicate the observed resistivity minima above $T_{N1}$.}
 \end{figure*}

 \begin{figure*}[ht]
    \centering
	\adjincludegraphics[height=7.5cm,trim={1cm 1cm 0.05cm 0.1cm},clip]{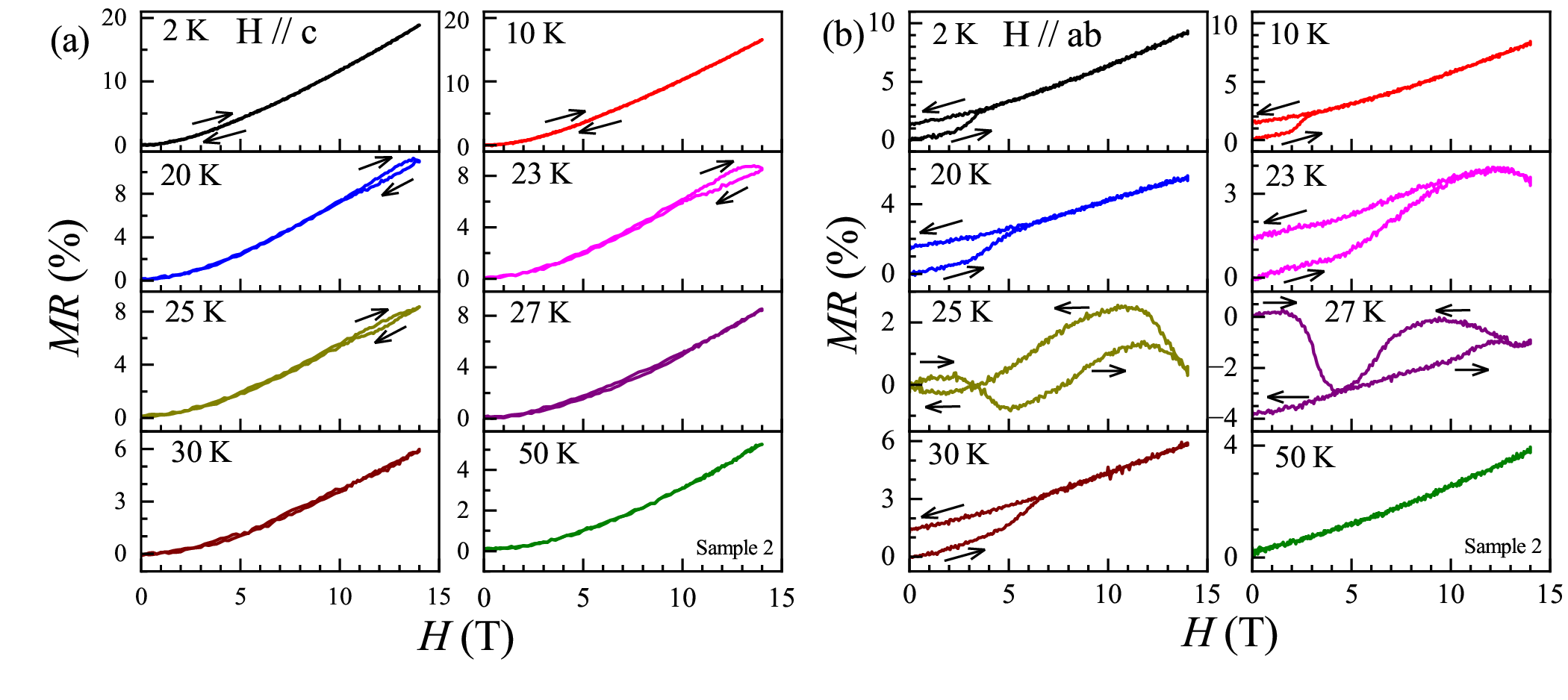}
    \caption{Magnetoresistance as a function of magnetic field (0–14 T, both up and down cycles) for different temperatures in the close vicinity of $T_{N1}$ and $T_{N2}$ for both $H \parallel c$ and $H \parallel ab$ directions as shown in (a) and (b), respectively. The arrows mark the way the applied magnetic field is varied.}
 \end{figure*}

To investigate the changes induced by the magnetic field, we conducted isothermal magnetization, M(H) measurements at selected temperatures below 50 K. Fig.~4 presents detailed M(H) curves measured at different temperatures for $H \parallel c$ (Fig.~4(a)) and $H \parallel ab$ (Fig.~4(b)). Interestingly, at 2 K there is a notable change in the slope of M(H) near 5 T, indicated by an arrow in the case of $H \parallel c$ as shown in Fig.4(a). This metamagnetic-like field-induced transition gradually gets smeared out with increasing temperature approaching $T_{N2}$. To capture this feature, we have plotted $\Delta M$ in the upper inset of Fig.~4(a), obtained by subtracting the linear component from M(H). On a larger field scale, the curves appear to be predominantly reversible with respect to field sweep direction. However, a zoom in the temperature range between two AFM transitions reveals a subtle hysteresis in M(H), shown in the lower inset of Fig.4(b). Interestingly, this hysteretic transition is absent for the $H \parallel ab$ orientation at the same field and temperature, but another hysteresis is observable at the lowest measured temperature of 2 K, as shown in Fig.4(b). The two insets in Fig.4(b) show the propagation of such hysteresis towards higher fields with an increase in the temperature. Emergence of such hysteresis at higher fields between two AFM transitions is further corroborated by features in the transport data (discussed below). Together, these direction, temperature, and field-dependent behaviors suggest a complex magnetic ground state in Gd$_{2}$AlSi$_{3}$. Moreover, M(H) curves do not show the tendency to saturate even in a 14 T field in the magnetically ordered state for both orientations. The magnetic moment at the lowest measured temperature of 2 K at 14 T is only ~3$\mu_{B}$/formula unit i.e., ~1.5$\mu_{B}$/Gd$^{3+}$, which is much less than the free Gd$^{3+}$ magnetic moment, an indication of a canted antiferromagnetic structure.

\subsection{Heat Capacity and Magnetocaloric Effect}

Fig. 5(a) presents zero-field and in-field measurements of the specific heat, C$_{P}(T)$, of Gd$_{2}$AlSi$_{3}$ as a function of temperature. Clearly, there is a well-defined $\lambda$ anomaly at $T_{N1}$= 32 K, and another peculiar magnetic transition is noted at $T_{N2}$= 23 K, which is also inferred from the $\chi$(T) data. In addition, there is a hump showing up near 9 K which is close to 25\% of its long-range ordering temperature $T_{N1}$. Such phenomenon was observed to be common in detailed studies of specific heat of Gd-based compounds \cite{PhysRevB.94.144434, PhysRevB.43.13137, PhysRevB.43.13145, KONG2014212} and was explained to arise naturally for $(2J+1)$-fold degenerate multiplets as predicted by mean field calculations. Both the peaks at $T_{N1}$ and $T_{N2}$ gradually shift towards lower temperatures with an increase in applied magnetic fields, which endorses that both the transitions in all cases are of AFM types. However, note that the two peak values behave differently, as one gets enhanced and the other one depressed with increasing applied magnetic field, particularly beyond 5 T. Such unusual behavior might be due to a modulated AFM structure that needs to be further investigated by a scattering experiment. 
The inset shows the magnetic entropy $S_{4f}$ estimated from the magnetic part ($4f$ contribution) of heat capacity $C_{mag}$ to the total. C$_{mag}$ is obtained by estimating the lattice contribution using the heat capacity of non-magnetic La$_{2}$AlSi$_{3}$ as a reference \cite {PhysRevB.43.13137}. At the magnetic transition temperature, $S_{4f} \simeq  12$ J/Gd-mol K is $\sim$ 70\% of $R ln 8$ as expected for complete removal of the (2J+1)-fold degeneracy of the CEF ground-state of Gd$^{3+}$ ion. This reduced entropy value might be due to the substantial modulated magnetic structure and/or the presence of short-range correlations above long-range magnetic ordering. Further, Fig. 5(b) presents the comparison of $C(T)/T$ and the derivative of electrical resistivity $d\rho/dT$, showing a similar temperature dependence, importantly including a feature near 9 K, which corroborates the (2J+1)-fold degenerate multiplets in such Gd system \cite{KONG2014212, PhysRevLett.20.665}. Also, note the enhanced value of the linear temperature coefficient $\gamma$ of 545 mJ/mol/K$^{2}$, as extrapolated from the paramagnetic state. However, this is also possible due to the extraction of gamma from much higher temperatures. It is also necessary to confirm this by having the exact information of the crystal field splitting below the magnetic ordering temperature in Gd$_{2}$AlSi$_{3}$. This is a surprisingly high value for a Gd-based system in which the $f$-orbital is mostly localized. Although more detailed experiments are required to verify such a high $\gamma$ value in Gd$_{2}$AlSi$_{3}$, this motivates further investigation of related Gd-based compounds to identify strongly correlated electron behavior \cite{PhysRevB.51.8631,PhysRevB.51.8178, PhysRevB.55.8369,PhysRevB.58.9178,Chandragiri_2016}.

We have also studied the magnetocaloric effect (MCE) in Gd$_{2}$AlSi$_{3}$ to get more information about the order of magnetic phase transitions, inspired by a previous study on Gd$_{5}$Si$_{2}$Ge$_{2}$ \cite{PhysRevLett.78.4494}. By measuring the isothermal magnetization at different temperatures with close intervals of 2 K up to 100 K, we can calculate the MCE using Maxwell's relations \cite {tishin2016magnetocaloric}. We have derived the isothermal entropy change, defined as -$\Delta S$ up to 100 K (i.e., to a much higher temperature than $T_N$), and the results obtained are shown in Fig. 6(a) and (b). Looking at the curves in Fig. 6(a), the sign of -$\Delta S$ is negative at low temperature, until about 10 K for fields up to 7 T, which implies the dominance of an AFM component \cite{PECHARSKY199944} over the FM phase to magnetocaloric effect; at higher temperatures, the sign of -$\Delta S$ is positive, which is a signature \cite{LI2020153810} of a tendency for spin-reorientation, suggesting the dominance of FM correlations in the presence of such magnitudes of external fields. Notably, a broad maximum starts to appear around $T_{N1}$. The value of -$\Delta$S increases further even in the paramagnetic state, near 50 K, and remains positive even up to 100 K (see Fig.~6b). Such a broad maximum over a wide temperature range well above $T_{N1}$ appears to support the proposal that FM clusters gradually form well above the long-range magnetic ordering. Interestingly, this is further substantiated by the observed transport properties in the paramagnetic range in such Gd-bases system. The collective results indicate that this compound could exemplify intriguing magnetic precursor effects in the paramagnetic state mainly in heavy rare-earths compounds.

\subsection{Magnetotransport Properties}

Motivated by recently reported exotic magnetotransport experiemnts confirming a skyrmion phase in Gd$_{2}$PdSi$_{3}$ \cite{kurumaji2019skyrmion}, we have measured temperature-dependent electrical resistivity $\rho(T)$ for both field directions, $H \parallel c$ and $H \parallel ab$ plane in Gd$_{2}$AlSi$_{3}$, as shown in Fig.~7(a) and (b) respectively. While the general trend of $\rho(T)$ in the paramagnetic state of Gd$_{2}$AlSi$_{3}$ is metallic on cooling below room temperature, anomalous behavior is observed on approaching $T_{N1}$ from higher temperature: 
on cooling, the resistivity incurs an upturn and sharp peak denoting the first AFM transition at $T_{N1}$. 
Similar behavior has been reported in other heavy rare-earth compounds, but over a wider temperature range \cite{PhysRevLett.91.036603,KUMAR2019165515,KUMAR2021168285}. In contrast, no resistivity minimum was observed in several Gd-based compounds (e.g., GdAg$_{2}$Si$_{2}$, GdAu$_{2}$Si$_{2}$, GdCu$_{2}$Ge$_{2}$, GdPd$_{2}$Ge$_{2}$\cite{PhysRevB.58.9178}, and Gd$_{4}$RhAl \cite{PhysRevMaterials.5.054407}), 
while there are exceptions in some families such as R$_{2}$RhSi$_{3}$ \cite{PhysRevB.101.144440}, and RCuAs$_{2}$\cite{PhysRevLett.91.036603}, R$_{4}$PtAl \cite{KUMAR2019165515, 10.1063/1.5121293, 10.1063/5.0016724,magnetochemistry9030085}. The origin of the abrupt upturn in Gd$_{2}$AlSi$_{3}$ just above $T_{N1}$ may be due to critical scattering on approach to the ordered phase, but such a conclusion will require further studies of the magnetic structure.

Below $T_{N1}$, the loss of spin-disorder scattering due to long-range AFM ordering leads to a rapid decrease in the resistivity, which remains featureless in the absence of an applied field. However, increasing magnetic field reveals a secondary feature near 23 K, which we attribute to the appearance of a second magnetic phase transition that is more prominent in data measured in the $H \parallel ab$ configuration, indicating anisotropy between crystallographic directions in the magnetically ordered state specifically between $T_{N1}$ and $T_{N2}$.\\
 \begin{figure}[h]
	\adjincludegraphics[height=14cm,trim={2.6cm 6cm 2.2cm 4cm},clip]{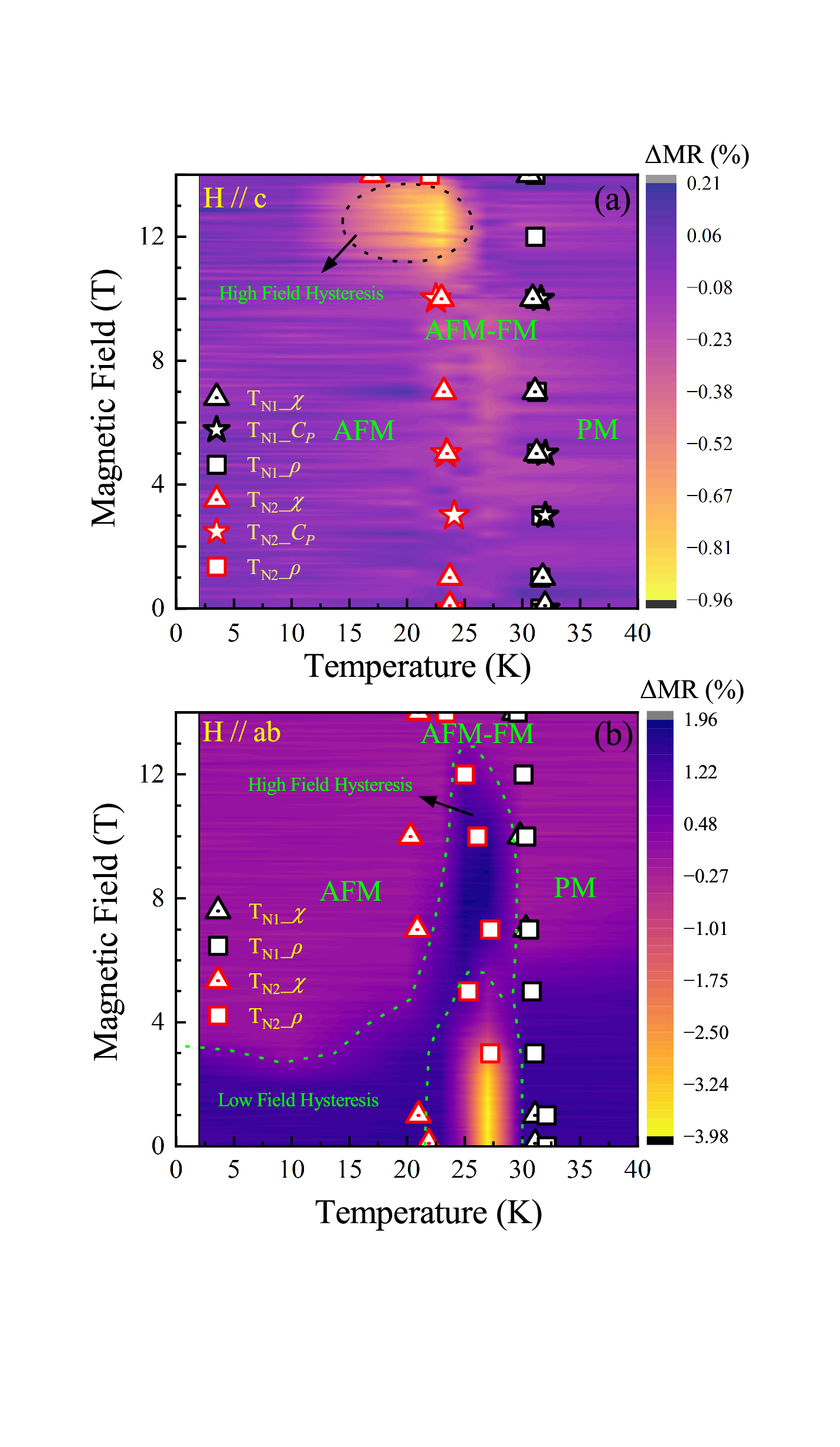}
    \caption{ Contour plot of $\Delta$MR (to present the existing hysteresis in MR) with the magnetic phase diagram of Gd$_{2}$AlSi$_{3}$ which depicts the existing magnetic phase transitions determined from magnetic susceptibility, heat capacity and electrical transport measurements performed under different applied magnetic fields. Points denoted by triangles and stars are determined from the d$\chi(T)$/dt and dC$_P$/dT curves and squares are obtained from $\rho(T)$ plots showing the antiferromagnetic transition/anomalies. The black symbols are $T_{N1}$ and red symbols are $T_{N2}$.}
    
 \end{figure}
 
 To address this observed complexity in $\rho(T)$, especially between $T_{N1}$ and $T_{N2}$, we have taken isothermal MR at  temperatures focused between $T_{N1}$ and $T_{N2}$, as shown in Fig. 8. In general, the MR behavior in the paramagnetic range looks similar in both directions, although the MR value in case of $H \parallel c$ is higher than $H \parallel ab$ direction. Interesting features, like hysteresis in MR, are observed at temperatures between the two transitions, although for $H \parallel ab$, there is a hysteretic behavior that extends down to 2~K.  As shown in Fig. 8(a), in the range 20-25 K the MR exhibits a clear hysteretic loop above 10~T that is absent at other temperatures, indicating a delicate competition between AFM and FM correlations. Fig. 8(b) depicts an even more exotic behavior for $H \parallel ab$, as a field-driven MR change from positive to negative occurs above 10~T. Such magnetic field-induced MR changes from positive to negative are generally associated with an AFM ground state \cite{10.1063/1.115758}. A careful look at the curves at 25 K suggests that the hysteretic loops along $H \parallel ab$ are wider in the magnetic field range than those along $H \parallel c$. The sign of MR is positive as expected for an ideal antiferromagnet, however, there is a negative MR contribution along $H \parallel ab$, which implies the existence of magnetic superzone gaps and/or the role of anisotropy as also seen in magnetism between the two magnetic transition temperatures. Such magnetic anisotropy might be possible due to the anisotropic exchange interactions arising from the Gd-triangular lattice. These observations further suggest the existence of a complex magnetic structure, which can be further elucidated via higher-field magnetotransport experiments.

Based on the experimental data presented above, we present field-temperature (\textit{B}-\textit{T}) phase diagrams for both magnetic field orientations to characterize the phase boundaries of the various features. The phase transition temperatures are determined from the peak position of derivatives of magnetic susceptibility, heat capacity, and the peak positions in transport measurements. As shown in Fig. 9, the peak positions of derivatives of the $\chi(T)$, C$_P$, $\rho(T)$, and MR data are used to identify boundaries, revealing two distinct regions. The first region below the magnetic order $T_{N1}$ and above $T_{N2}$, labeled AFM-FM, corresponds to a competitive phase between the AFM and FM ordering, which turns out to be the most complex phase. The second region labeled AFM appears below $T_{N2}$. The unusual high-field hysteresis in MR observed for both applied field directions presents the most exotic features in the resulting phase diagram. Furthermore, a low-field MR-hysteresis is also present along with a high-field MR-hysteresis for $H \parallel ab$ as shown in Fig. 9(b). As previously discussed, there is a temperature range between the two transitions where anomalies persist, which are possibly due to a non-collinear magnetic phase in the AFM-FM region. Such a multi-phase structure is reminiscent of skyrmionic phases that arise due to competing energy scales with comparable magnitudes \cite{doi:10.1021/acs.chemrev.0c00297}. The present findings are consistent with skyrmion-hosting systems, however, future magnetic structure studies are needed to confirm this hypothesis, as done in similar centrosymmetric materials \cite{PhysRevB.108.L100404, Yasui2020}. 

\section{Conclusion}
In conclusion, we present a comprehensive study of the magnetic and transport properties of newly synthesized single-crystals of Gd$_{2}$AlSi$_{3}$, which crystallize in the centrosymmetric tetragonal $I4_{1}/amd$ structure and contain a triangular arrangement of single-site Gd atoms. The magnetic and transport behavior of this system is quite complex, with a fascinating magnetic anisotropy that materializes as different temperature and field-dependent hysteresis regimes observed along different magnetic field directions. With two magnetic transitions, signatures of competition between antiferromagnetic and ferromagnetic correlations, strong anisotropic features, and complex hysteresis in magnetoresistance, this system presents a plethora of interesting features that may harbor signatures of skyrmionic magnetism akin to other related compounds. The recent discovery of Gd-based skyrmion host materials \cite{kurumaji2019skyrmion,Hirschberger2019-dw,Khanh2020-ra} emphasizes the crucial role of the rare earth cation in the distinctive properties found in Gd$_{2}$AlSi$_{3}$. Future studies of this series of aluminum-based 213 materials will be of interest to further explore the propensity toward skyrmion phases in centrosymmetric systems.

\section{acknowledgments}

Research at the University of Maryland was supported by the Gordon and Betty Moore Foundation’s EPiQS Initiative through Grant No. GBMF9071, the U.S. National Science Foundation (NSF) Grant No. DMR2303090, the Binational Science Foundation Grant No. 2022126, and the Maryland Quantum Materials Center.
R.K. acknowledges support from the NSF Grant Number 2201516 under the Accelnet program of Office of International Science and Engineering (OISE).
S.R.S. acknowledges support from the National Institute of Standards and Technology Cooperative Agreement 70NANB17H301. We also thank Prof K. Maiti (TIFR, Mumbai) for providing experimental facilities use.


\nocite{*}

\bibliography{Gd2AlSi3}

\begin{thebibliography}{57}
\expandafter\ifx\csname natexlab\endcsname\relax\def\natexlab#1{#1}\fi
\expandafter\ifx\csname bibnamefont\endcsname\relax
  \def\bibnamefont#1{#1}\fi
\expandafter\ifx\csname bibfnamefont\endcsname\relax
  \def\bibfnamefont#1{#1}\fi
\expandafter\ifx\csname citenamefont\endcsname\relax
  \def\citenamefont#1{#1}\fi
\expandafter\ifx\csname url\endcsname\relax
  \def\url#1{\texttt{#1}}\fi
\expandafter\ifx\csname urlprefix\endcsname\relax\def\urlprefix{URL }\fi
\providecommand{\bibinfo}[2]{#2}
\providecommand{\eprint}[2][]{\url{#2}}

\bibitem[{\citenamefont{Batista et~al.}(2016)\citenamefont{Batista, Lin, Hayami, and Kamiya}}]{Batista_2016}
\bibinfo{author}{\bibfnamefont{C.~D.} \bibnamefont{Batista}}, \bibinfo{author}{\bibfnamefont{S.-Z.} \bibnamefont{Lin}}, \bibinfo{author}{\bibfnamefont{S.}~\bibnamefont{Hayami}}, \bibnamefont{and} \bibinfo{author}{\bibfnamefont{Y.}~\bibnamefont{Kamiya}}, \bibinfo{journal}{Reports on Progress in Physics} \textbf{\bibinfo{volume}{79}}, \bibinfo{pages}{084504} (\bibinfo{year}{2016}), \urlprefix\url{https://dx.doi.org/10.1088/0034-4885/79/8/084504}.

\bibitem[{\citenamefont{Xia et~al.}(2019)\citenamefont{Xia, Zhang, Ezawa, Hou, Wang, Liu, and Zhou}}]{PhysRevApplied.11.044046}
\bibinfo{author}{\bibfnamefont{J.}~\bibnamefont{Xia}}, \bibinfo{author}{\bibfnamefont{X.}~\bibnamefont{Zhang}}, \bibinfo{author}{\bibfnamefont{M.}~\bibnamefont{Ezawa}}, \bibinfo{author}{\bibfnamefont{Z.}~\bibnamefont{Hou}}, \bibinfo{author}{\bibfnamefont{W.}~\bibnamefont{Wang}}, \bibinfo{author}{\bibfnamefont{X.}~\bibnamefont{Liu}}, \bibnamefont{and} \bibinfo{author}{\bibfnamefont{Y.}~\bibnamefont{Zhou}}, \bibinfo{journal}{Phys. Rev. Appl.} \textbf{\bibinfo{volume}{11}}, \bibinfo{pages}{044046} (\bibinfo{year}{2019}), \urlprefix\url{https://link.aps.org/doi/10.1103/PhysRevApplied.11.044046}.

\bibitem[{\citenamefont{Wang et~al.}(2016)\citenamefont{Wang, Barros, Chern, Maslov, and Batista}}]{PhysRevLett.117.206601}
\bibinfo{author}{\bibfnamefont{Z.}~\bibnamefont{Wang}}, \bibinfo{author}{\bibfnamefont{K.}~\bibnamefont{Barros}}, \bibinfo{author}{\bibfnamefont{G.-W.} \bibnamefont{Chern}}, \bibinfo{author}{\bibfnamefont{D.~L.} \bibnamefont{Maslov}}, \bibnamefont{and} \bibinfo{author}{\bibfnamefont{C.~D.} \bibnamefont{Batista}}, \bibinfo{journal}{Phys. Rev. Lett.} \textbf{\bibinfo{volume}{117}}, \bibinfo{pages}{206601} (\bibinfo{year}{2016}), \urlprefix\url{https://link.aps.org/doi/10.1103/PhysRevLett.117.206601}.

\bibitem[{\citenamefont{Wang and Batista}(2020)}]{PhysRevB.101.184432}
\bibinfo{author}{\bibfnamefont{Z.}~\bibnamefont{Wang}} \bibnamefont{and} \bibinfo{author}{\bibfnamefont{C.~D.} \bibnamefont{Batista}}, \bibinfo{journal}{Phys. Rev. B} \textbf{\bibinfo{volume}{101}}, \bibinfo{pages}{184432} (\bibinfo{year}{2020}), \urlprefix\url{https://link.aps.org/doi/10.1103/PhysRevB.101.184432}.

\bibitem[{\citenamefont{Mekata}(1977)}]{doi:10.1143/JPSJ.42.76}
\bibinfo{author}{\bibfnamefont{M.}~\bibnamefont{Mekata}}, \bibinfo{journal}{Journal of the Physical Society of Japan} \textbf{\bibinfo{volume}{42}}, \bibinfo{pages}{76} (\bibinfo{year}{1977}).

\bibitem[{\citenamefont{Yelon et~al.}(1975)\citenamefont{Yelon, Cox, and Eibsch\"utz}}]{PhysRevB.12.5007}
\bibinfo{author}{\bibfnamefont{W.~B.} \bibnamefont{Yelon}}, \bibinfo{author}{\bibfnamefont{D.~E.} \bibnamefont{Cox}}, \bibnamefont{and} \bibinfo{author}{\bibfnamefont{M.}~\bibnamefont{Eibsch\"utz}}, \bibinfo{journal}{Phys. Rev. B} \textbf{\bibinfo{volume}{12}}, \bibinfo{pages}{5007} (\bibinfo{year}{1975}), \urlprefix\url{https://link.aps.org/doi/10.1103/PhysRevB.12.5007}.

\bibitem[{\citenamefont{Sampathkumaran and Niazi}(2002)}]{PhysRevB.65.180401}
\bibinfo{author}{\bibfnamefont{E.~V.} \bibnamefont{Sampathkumaran}} \bibnamefont{and} \bibinfo{author}{\bibfnamefont{A.}~\bibnamefont{Niazi}}, \bibinfo{journal}{Phys. Rev. B} \textbf{\bibinfo{volume}{65}}, \bibinfo{pages}{180401} (\bibinfo{year}{2002}), \urlprefix\url{https://link.aps.org/doi/10.1103/PhysRevB.65.180401}.

\bibitem[{\citenamefont{Shaginyan et~al.}(2020)\citenamefont{Shaginyan, Stephanovich, Msezane, Japaridze, Clark, Amusia, and Kirichenko}}]{Shaginyan}
\bibinfo{author}{\bibfnamefont{V.~R.} \bibnamefont{Shaginyan}}, \bibinfo{author}{\bibfnamefont{V.~A.} \bibnamefont{Stephanovich}}, \bibinfo{author}{\bibfnamefont{A.~Z.} \bibnamefont{Msezane}}, \bibinfo{author}{\bibfnamefont{G.~S.} \bibnamefont{Japaridze}}, \bibinfo{author}{\bibfnamefont{J.~W.} \bibnamefont{Clark}}, \bibinfo{author}{\bibfnamefont{M.~Y.} \bibnamefont{Amusia}}, \bibnamefont{and} \bibinfo{author}{\bibfnamefont{E.~V.} \bibnamefont{Kirichenko}}, \bibinfo{journal}{J. Mater. Sci.} \textbf{\bibinfo{volume}{55}}, \bibinfo{pages}{2257} (\bibinfo{year}{2020}), \urlprefix\url{https://doi.org/10.1007/s10853-019-04128-w}.

\bibitem[{\citenamefont{Levy et~al.}(2008)\citenamefont{Levy, Sheikin, Berthier, Horvatić, Takigawa, Kageyama, Waki, and Ueda}}]{Levy_2008}
\bibinfo{author}{\bibfnamefont{F.}~\bibnamefont{Levy}}, \bibinfo{author}{\bibfnamefont{I.}~\bibnamefont{Sheikin}}, \bibinfo{author}{\bibfnamefont{C.}~\bibnamefont{Berthier}}, \bibinfo{author}{\bibfnamefont{M.}~\bibnamefont{Horvatić}}, \bibinfo{author}{\bibfnamefont{M.}~\bibnamefont{Takigawa}}, \bibinfo{author}{\bibfnamefont{H.}~\bibnamefont{Kageyama}}, \bibinfo{author}{\bibfnamefont{T.}~\bibnamefont{Waki}}, \bibnamefont{and} \bibinfo{author}{\bibfnamefont{Y.}~\bibnamefont{Ueda}}, \bibinfo{journal}{Europhysics Letters} \textbf{\bibinfo{volume}{81}}, \bibinfo{pages}{67004} (\bibinfo{year}{2008}), \urlprefix\url{https://dx.doi.org/10.1209/0295-5075/81/67004}.

\bibitem[{\citenamefont{Bezvershenko et~al.}(2018)\citenamefont{Bezvershenko, Kolezhuk, and Ivanov}}]{PhysRevB.97.054408}
\bibinfo{author}{\bibfnamefont{A.~V.} \bibnamefont{Bezvershenko}}, \bibinfo{author}{\bibfnamefont{A.~K.} \bibnamefont{Kolezhuk}}, \bibnamefont{and} \bibinfo{author}{\bibfnamefont{B.~A.} \bibnamefont{Ivanov}}, \bibinfo{journal}{Phys. Rev. B} \textbf{\bibinfo{volume}{97}}, \bibinfo{pages}{054408} (\bibinfo{year}{2018}), \urlprefix\url{https://link.aps.org/doi/10.1103/PhysRevB.97.054408}.

\bibitem[{\citenamefont{Gordon et~al.}(1997)\citenamefont{Gordon, Warren, Alexander, DiSalvo, and P{\"o}ttgen}}]{gordon1997substitution}
\bibinfo{author}{\bibfnamefont{R.}~\bibnamefont{Gordon}}, \bibinfo{author}{\bibfnamefont{C.}~\bibnamefont{Warren}}, \bibinfo{author}{\bibfnamefont{M.}~\bibnamefont{Alexander}}, \bibinfo{author}{\bibfnamefont{F.}~\bibnamefont{DiSalvo}}, \bibnamefont{and} \bibinfo{author}{\bibfnamefont{R.}~\bibnamefont{P{\"o}ttgen}}, \bibinfo{journal}{Journal of alloys and compounds} \textbf{\bibinfo{volume}{248}}, \bibinfo{pages}{24} (\bibinfo{year}{1997}).

\bibitem[{\citenamefont{Saha et~al.}(1999)\citenamefont{Saha, Sugawara, Matsuda, Sato, Mallik, and Sampathkumaran}}]{10.1103/physrevb.60.12162}
\bibinfo{author}{\bibfnamefont{S.}~\bibnamefont{Saha}}, \bibinfo{author}{\bibfnamefont{H.}~\bibnamefont{Sugawara}}, \bibinfo{author}{\bibfnamefont{T.~D.} \bibnamefont{Matsuda}}, \bibinfo{author}{\bibfnamefont{H.}~\bibnamefont{Sato}}, \bibinfo{author}{\bibfnamefont{R.~C.} \bibnamefont{Mallik}}, \bibnamefont{and} \bibinfo{author}{\bibfnamefont{E.~V.} \bibnamefont{Sampathkumaran}}, \bibinfo{journal}{Physical Review B} \textbf{\bibinfo{volume}{60}}, \bibinfo{pages}{12162} (\bibinfo{year}{1999}).

\bibitem[{\citenamefont{Tien et~al.}(2000)\citenamefont{Tien, Feng, Wur, and Lu}}]{tien20002}
\bibinfo{author}{\bibfnamefont{C.}~\bibnamefont{Tien}}, \bibinfo{author}{\bibfnamefont{C.~H.} \bibnamefont{Feng}}, \bibinfo{author}{\bibfnamefont{C.~S.} \bibnamefont{Wur}}, \bibnamefont{and} \bibinfo{author}{\bibfnamefont{J.~J.} \bibnamefont{Lu}}, \bibinfo{journal}{Physical Review B} \textbf{\bibinfo{volume}{61}}, \bibinfo{pages}{12151} (\bibinfo{year}{2000}).

\bibitem[{\citenamefont{Majumdar et~al.}(2001)\citenamefont{Majumdar, Bitterlich, Behr, L{\"o}ser, Paulose, and Sampathkumaran}}]{majumdar2001magnetic}
\bibinfo{author}{\bibfnamefont{S.}~\bibnamefont{Majumdar}}, \bibinfo{author}{\bibfnamefont{H.}~\bibnamefont{Bitterlich}}, \bibinfo{author}{\bibfnamefont{G.}~\bibnamefont{Behr}}, \bibinfo{author}{\bibfnamefont{W.}~\bibnamefont{L{\"o}ser}}, \bibinfo{author}{\bibfnamefont{P.}~\bibnamefont{Paulose}}, \bibnamefont{and} \bibinfo{author}{\bibfnamefont{E.}~\bibnamefont{Sampathkumaran}}, \bibinfo{journal}{Physical Review B} \textbf{\bibinfo{volume}{64}}, \bibinfo{pages}{012418} (\bibinfo{year}{2001}).

\bibitem[{\citenamefont{Szlawska et~al.}(2010)\citenamefont{Szlawska, Kaczorowski, and Reehuis}}]{szlawska2010experimental}
\bibinfo{author}{\bibfnamefont{M.}~\bibnamefont{Szlawska}}, \bibinfo{author}{\bibfnamefont{D.}~\bibnamefont{Kaczorowski}}, \bibnamefont{and} \bibinfo{author}{\bibfnamefont{M.}~\bibnamefont{Reehuis}}, \bibinfo{journal}{Physical Review B} \textbf{\bibinfo{volume}{81}}, \bibinfo{pages}{094423} (\bibinfo{year}{2010}).

\bibitem[{\citenamefont{Tang et~al.}(2011)\citenamefont{Tang, Frontzek, Dshemuchadse, Leisegang, Zschornak, Mietrach, Hoffmann, L{\"o}ser, Gemming, Meyer et~al.}}]{tang2011crystallographic}
\bibinfo{author}{\bibfnamefont{F.}~\bibnamefont{Tang}}, \bibinfo{author}{\bibfnamefont{M.}~\bibnamefont{Frontzek}}, \bibinfo{author}{\bibfnamefont{J.}~\bibnamefont{Dshemuchadse}}, \bibinfo{author}{\bibfnamefont{T.}~\bibnamefont{Leisegang}}, \bibinfo{author}{\bibfnamefont{M.}~\bibnamefont{Zschornak}}, \bibinfo{author}{\bibfnamefont{R.}~\bibnamefont{Mietrach}}, \bibinfo{author}{\bibfnamefont{J.-U.} \bibnamefont{Hoffmann}}, \bibinfo{author}{\bibfnamefont{W.}~\bibnamefont{L{\"o}ser}}, \bibinfo{author}{\bibfnamefont{S.}~\bibnamefont{Gemming}}, \bibinfo{author}{\bibfnamefont{D.~C.} \bibnamefont{Meyer}}, \bibnamefont{et~al.}, \bibinfo{journal}{Physical Review B} \textbf{\bibinfo{volume}{84}}, \bibinfo{pages}{104105} (\bibinfo{year}{2011}).

\bibitem[{\citenamefont{{Baumbach} et~al.}(2015)\citenamefont{{Baumbach}, {Gallagher}, {Besara}, {Sun}, {Siegrist}, {Singh}, {Thompson}, {Ronning}, and {Bauer}}}]{2015PhRvB..91c5102B}
\bibinfo{author}{\bibfnamefont{R.~E.} \bibnamefont{{Baumbach}}}, \bibinfo{author}{\bibfnamefont{A.}~\bibnamefont{{Gallagher}}}, \bibinfo{author}{\bibfnamefont{T.}~\bibnamefont{{Besara}}}, \bibinfo{author}{\bibfnamefont{J.}~\bibnamefont{{Sun}}}, \bibinfo{author}{\bibfnamefont{T.}~\bibnamefont{{Siegrist}}}, \bibinfo{author}{\bibfnamefont{D.~J.} \bibnamefont{{Singh}}}, \bibinfo{author}{\bibfnamefont{J.~D.} \bibnamefont{{Thompson}}}, \bibinfo{author}{\bibfnamefont{F.}~\bibnamefont{{Ronning}}}, \bibnamefont{and} \bibinfo{author}{\bibfnamefont{E.~D.} \bibnamefont{{Bauer}}}, \bibinfo{journal}{\prb} \textbf{\bibinfo{volume}{91}}, \bibinfo{eid}{035102} (\bibinfo{year}{2015}).

\bibitem[{\citenamefont{Pakhira et~al.}(2016)\citenamefont{Pakhira, Mazumdar, Ranganathan, Giri, and Avdeev}}]{pakhira2016large}
\bibinfo{author}{\bibfnamefont{S.}~\bibnamefont{Pakhira}}, \bibinfo{author}{\bibfnamefont{C.}~\bibnamefont{Mazumdar}}, \bibinfo{author}{\bibfnamefont{R.}~\bibnamefont{Ranganathan}}, \bibinfo{author}{\bibfnamefont{S.}~\bibnamefont{Giri}}, \bibnamefont{and} \bibinfo{author}{\bibfnamefont{M.}~\bibnamefont{Avdeev}}, \bibinfo{journal}{Physical Review B} \textbf{\bibinfo{volume}{94}}, \bibinfo{pages}{104414} (\bibinfo{year}{2016}).

\bibitem[{\citenamefont{Kurumaji et~al.}(2019)\citenamefont{Kurumaji, Nakajima, Hirschberger, Kikkawa, Yamasaki, Sagayama, Nakao, Taguchi, Arima, and Tokura}}]{kurumaji2019skyrmion}
\bibinfo{author}{\bibfnamefont{T.}~\bibnamefont{Kurumaji}}, \bibinfo{author}{\bibfnamefont{T.}~\bibnamefont{Nakajima}}, \bibinfo{author}{\bibfnamefont{M.}~\bibnamefont{Hirschberger}}, \bibinfo{author}{\bibfnamefont{A.}~\bibnamefont{Kikkawa}}, \bibinfo{author}{\bibfnamefont{Y.}~\bibnamefont{Yamasaki}}, \bibinfo{author}{\bibfnamefont{H.}~\bibnamefont{Sagayama}}, \bibinfo{author}{\bibfnamefont{H.}~\bibnamefont{Nakao}}, \bibinfo{author}{\bibfnamefont{Y.}~\bibnamefont{Taguchi}}, \bibinfo{author}{\bibfnamefont{T.-h.} \bibnamefont{Arima}}, \bibnamefont{and} \bibinfo{author}{\bibfnamefont{Y.}~\bibnamefont{Tokura}}, \bibinfo{journal}{Science} \textbf{\bibinfo{volume}{365}}, \bibinfo{pages}{914} (\bibinfo{year}{2019}).

\bibitem[{\citenamefont{Kumar et~al.}(2020{\natexlab{a}})\citenamefont{Kumar, Iyer, Paulose, and Sampathkumaran}}]{PhysRevB.101.144440}
\bibinfo{author}{\bibfnamefont{R.}~\bibnamefont{Kumar}}, \bibinfo{author}{\bibfnamefont{K.~K.} \bibnamefont{Iyer}}, \bibinfo{author}{\bibfnamefont{P.~L.} \bibnamefont{Paulose}}, \bibnamefont{and} \bibinfo{author}{\bibfnamefont{E.~V.} \bibnamefont{Sampathkumaran}}, \bibinfo{journal}{Phys. Rev. B} \textbf{\bibinfo{volume}{101}}, \bibinfo{pages}{144440} (\bibinfo{year}{2020}{\natexlab{a}}), \urlprefix\url{https://link.aps.org/doi/10.1103/PhysRevB.101.144440}.

\bibitem[{\citenamefont{Szytu{\l}a et~al.}(1999)\citenamefont{Szytu{\l}a, Hofmann, Penc, {\'S}laski, Majumdar, Sampathkumaran, and Zygmunt}}]{szytula1999magnetic}
\bibinfo{author}{\bibfnamefont{A.}~\bibnamefont{Szytu{\l}a}}, \bibinfo{author}{\bibfnamefont{M.}~\bibnamefont{Hofmann}}, \bibinfo{author}{\bibfnamefont{B.}~\bibnamefont{Penc}}, \bibinfo{author}{\bibfnamefont{M.}~\bibnamefont{{\'S}laski}}, \bibinfo{author}{\bibfnamefont{S.}~\bibnamefont{Majumdar}}, \bibinfo{author}{\bibfnamefont{E.}~\bibnamefont{Sampathkumaran}}, \bibnamefont{and} \bibinfo{author}{\bibfnamefont{A.}~\bibnamefont{Zygmunt}}, \bibinfo{journal}{Journal of magnetism and magnetic materials} \textbf{\bibinfo{volume}{202}}, \bibinfo{pages}{365} (\bibinfo{year}{1999}).

\bibitem[{\citenamefont{Li et~al.}(2001)\citenamefont{Li, Nimori, Shiokawa, Haga, Yamamoto, and Onuki}}]{li2001magnetic}
\bibinfo{author}{\bibfnamefont{D.}~\bibnamefont{Li}}, \bibinfo{author}{\bibfnamefont{S.}~\bibnamefont{Nimori}}, \bibinfo{author}{\bibfnamefont{Y.}~\bibnamefont{Shiokawa}}, \bibinfo{author}{\bibfnamefont{Y.}~\bibnamefont{Haga}}, \bibinfo{author}{\bibfnamefont{E.}~\bibnamefont{Yamamoto}}, \bibnamefont{and} \bibinfo{author}{\bibfnamefont{Y.}~\bibnamefont{Onuki}}, \bibinfo{journal}{Solid state communications} \textbf{\bibinfo{volume}{120}}, \bibinfo{pages}{227} (\bibinfo{year}{2001}).

\bibitem[{\citenamefont{Smidman et~al.}(2019)\citenamefont{Smidman, Ritter, Adroja, Rayaprol, Basu, Sampathkumaran, and Hillier}}]{smidman2019magnetic}
\bibinfo{author}{\bibfnamefont{M.}~\bibnamefont{Smidman}}, \bibinfo{author}{\bibfnamefont{C.}~\bibnamefont{Ritter}}, \bibinfo{author}{\bibfnamefont{D.}~\bibnamefont{Adroja}}, \bibinfo{author}{\bibfnamefont{S.}~\bibnamefont{Rayaprol}}, \bibinfo{author}{\bibfnamefont{T.}~\bibnamefont{Basu}}, \bibinfo{author}{\bibfnamefont{E.}~\bibnamefont{Sampathkumaran}}, \bibnamefont{and} \bibinfo{author}{\bibfnamefont{A.}~\bibnamefont{Hillier}}, \bibinfo{journal}{Physical Review B} \textbf{\bibinfo{volume}{100}}, \bibinfo{pages}{134423} (\bibinfo{year}{2019}).

\bibitem[{\citenamefont{Litzbarski et~al.}(2020)\citenamefont{Litzbarski, Klimczuk, and Winiarski}}]{litzbarski2020synthesis}
\bibinfo{author}{\bibfnamefont{L.}~\bibnamefont{Litzbarski}}, \bibinfo{author}{\bibfnamefont{T.}~\bibnamefont{Klimczuk}}, \bibnamefont{and} \bibinfo{author}{\bibfnamefont{M.~J.} \bibnamefont{Winiarski}}, \bibinfo{journal}{Journal of Physics: Condensed Matter} \textbf{\bibinfo{volume}{32}}, \bibinfo{pages}{225706} (\bibinfo{year}{2020}).

\bibitem[{\citenamefont{Pan et~al.}(2013)\citenamefont{Pan, Cao, Bai, Song, Zheng, and Duan}}]{pan2013structures}
\bibinfo{author}{\bibfnamefont{Z.-Y.} \bibnamefont{Pan}}, \bibinfo{author}{\bibfnamefont{C.-D.} \bibnamefont{Cao}}, \bibinfo{author}{\bibfnamefont{X.-J.} \bibnamefont{Bai}}, \bibinfo{author}{\bibfnamefont{R.-B.} \bibnamefont{Song}}, \bibinfo{author}{\bibfnamefont{J.-B.} \bibnamefont{Zheng}}, \bibnamefont{and} \bibinfo{author}{\bibfnamefont{L.-B.} \bibnamefont{Duan}}, \bibinfo{journal}{Chinese Physics B} \textbf{\bibinfo{volume}{22}}, \bibinfo{pages}{056102} (\bibinfo{year}{2013}).

\bibitem[{\citenamefont{Mallik et~al.}(1998)\citenamefont{Mallik, Sampathkumaran, Strecker, and Wortmann}}]{R.Mallik_1998}
\bibinfo{author}{\bibfnamefont{R.}~\bibnamefont{Mallik}}, \bibinfo{author}{\bibfnamefont{E.~V.} \bibnamefont{Sampathkumaran}}, \bibinfo{author}{\bibfnamefont{M.}~\bibnamefont{Strecker}}, \bibnamefont{and} \bibinfo{author}{\bibfnamefont{G.}~\bibnamefont{Wortmann}}, \bibinfo{journal}{Europhysics Letters} \textbf{\bibinfo{volume}{41}}, \bibinfo{pages}{315} (\bibinfo{year}{1998}), \urlprefix\url{https://dx.doi.org/10.1209/epl/i1998-00149-4}.

\bibitem[{\citenamefont{Sampathkumaran}(2019)}]{sampathkumaran2019report}
\bibinfo{author}{\bibfnamefont{E.~V.} \bibnamefont{Sampathkumaran}}, \emph{\bibinfo{title}{A report of (topological) hall anomaly two decades ago in gd2pdsi3, and its relevance to the history of the field of topological hall effect due to magnetic skyrmions}} (\bibinfo{year}{2019}), \eprint{1910.09194}.

\bibitem[{\citenamefont{Paddison et~al.}(2022)\citenamefont{Paddison, Rai, May, Calder, Stone, Frontzek, and Christianson}}]{PhysRevLett.129.137202}
\bibinfo{author}{\bibfnamefont{J.~A.~M.} \bibnamefont{Paddison}}, \bibinfo{author}{\bibfnamefont{B.~K.} \bibnamefont{Rai}}, \bibinfo{author}{\bibfnamefont{A.~F.} \bibnamefont{May}}, \bibinfo{author}{\bibfnamefont{S.}~\bibnamefont{Calder}}, \bibinfo{author}{\bibfnamefont{M.~B.} \bibnamefont{Stone}}, \bibinfo{author}{\bibfnamefont{M.~D.} \bibnamefont{Frontzek}}, \bibnamefont{and} \bibinfo{author}{\bibfnamefont{A.~D.} \bibnamefont{Christianson}}, \bibinfo{journal}{Phys. Rev. Lett.} \textbf{\bibinfo{volume}{129}}, \bibinfo{pages}{137202} (\bibinfo{year}{2022}), \urlprefix\url{https://link.aps.org/doi/10.1103/PhysRevLett.129.137202}.

\bibitem[{\citenamefont{Sahu et~al.}(2021)\citenamefont{Sahu, Fobasso, and Strydom}}]{SAHU2021107214}
\bibinfo{author}{\bibfnamefont{B.}~\bibnamefont{Sahu}}, \bibinfo{author}{\bibfnamefont{R.~D.} \bibnamefont{Fobasso}}, \bibnamefont{and} \bibinfo{author}{\bibfnamefont{A.~M.} \bibnamefont{Strydom}}, \bibinfo{journal}{Intermetallics} \textbf{\bibinfo{volume}{135}}, \bibinfo{pages}{107214} (\bibinfo{year}{2021}), ISSN \bibinfo{issn}{0966-9795}, \urlprefix\url{https://www.sciencedirect.com/science/article/pii/S0966979521001308}.

\bibitem[{\citenamefont{Rietveld}(1969)}]{Rietveld_a07067}
\bibinfo{author}{\bibfnamefont{H.~M.} \bibnamefont{Rietveld}}, \bibinfo{journal}{Journal of Applied Crystallography} \textbf{\bibinfo{volume}{2}}, \bibinfo{pages}{65} (\bibinfo{year}{1969}), \urlprefix\url{https://doi.org/10.1107/S0021889869006558}.

\bibitem[{\citenamefont{Kong et~al.}(2016)\citenamefont{Kong, Meier, Lin, Saunders, Bud'ko, Flint, and Canfield}}]{PhysRevB.94.144434}
\bibinfo{author}{\bibfnamefont{T.}~\bibnamefont{Kong}}, \bibinfo{author}{\bibfnamefont{W.~R.} \bibnamefont{Meier}}, \bibinfo{author}{\bibfnamefont{Q.}~\bibnamefont{Lin}}, \bibinfo{author}{\bibfnamefont{S.~M.} \bibnamefont{Saunders}}, \bibinfo{author}{\bibfnamefont{S.~L.} \bibnamefont{Bud'ko}}, \bibinfo{author}{\bibfnamefont{R.}~\bibnamefont{Flint}}, \bibnamefont{and} \bibinfo{author}{\bibfnamefont{P.~C.} \bibnamefont{Canfield}}, \bibinfo{journal}{Phys. Rev. B} \textbf{\bibinfo{volume}{94}}, \bibinfo{pages}{144434} (\bibinfo{year}{2016}), \urlprefix\url{https://link.aps.org/doi/10.1103/PhysRevB.94.144434}.

\bibitem[{\citenamefont{Bouvier et~al.}(1991)\citenamefont{Bouvier, Lethuillier, and Schmitt}}]{PhysRevB.43.13137}
\bibinfo{author}{\bibfnamefont{M.}~\bibnamefont{Bouvier}}, \bibinfo{author}{\bibfnamefont{P.}~\bibnamefont{Lethuillier}}, \bibnamefont{and} \bibinfo{author}{\bibfnamefont{D.}~\bibnamefont{Schmitt}}, \bibinfo{journal}{Phys. Rev. B} \textbf{\bibinfo{volume}{43}}, \bibinfo{pages}{13137} (\bibinfo{year}{1991}), \urlprefix\url{https://link.aps.org/doi/10.1103/PhysRevB.43.13137}.

\bibitem[{\citenamefont{Blanco et~al.}(1991)\citenamefont{Blanco, Gignoux, and Schmitt}}]{PhysRevB.43.13145}
\bibinfo{author}{\bibfnamefont{J.~A.} \bibnamefont{Blanco}}, \bibinfo{author}{\bibfnamefont{D.}~\bibnamefont{Gignoux}}, \bibnamefont{and} \bibinfo{author}{\bibfnamefont{D.}~\bibnamefont{Schmitt}}, \bibinfo{journal}{Phys. Rev. B} \textbf{\bibinfo{volume}{43}}, \bibinfo{pages}{13145} (\bibinfo{year}{1991}), \urlprefix\url{https://link.aps.org/doi/10.1103/PhysRevB.43.13145}.

\bibitem[{\citenamefont{Kong et~al.}(2014)\citenamefont{Kong, Cunningham, Taufour, Budko, Buffon, Lin, Emmons, and Canfield}}]{KONG2014212}
\bibinfo{author}{\bibfnamefont{T.}~\bibnamefont{Kong}}, \bibinfo{author}{\bibfnamefont{C.~E.} \bibnamefont{Cunningham}}, \bibinfo{author}{\bibfnamefont{V.}~\bibnamefont{Taufour}}, \bibinfo{author}{\bibfnamefont{S.~L.} \bibnamefont{Budko}}, \bibinfo{author}{\bibfnamefont{M.~L.} \bibnamefont{Buffon}}, \bibinfo{author}{\bibfnamefont{X.}~\bibnamefont{Lin}}, \bibinfo{author}{\bibfnamefont{H.}~\bibnamefont{Emmons}}, \bibnamefont{and} \bibinfo{author}{\bibfnamefont{P.~C.} \bibnamefont{Canfield}}, \bibinfo{journal}{Journal of Magnetism and Magnetic Materials} \textbf{\bibinfo{volume}{358-359}}, \bibinfo{pages}{212} (\bibinfo{year}{2014}), ISSN \bibinfo{issn}{0304-8853}.

\bibitem[{\citenamefont{Fisher and Langer}(1968)}]{PhysRevLett.20.665}
\bibinfo{author}{\bibfnamefont{M.~E.} \bibnamefont{Fisher}} \bibnamefont{and} \bibinfo{author}{\bibfnamefont{J.~S.} \bibnamefont{Langer}}, \bibinfo{journal}{Phys. Rev. Lett.} \textbf{\bibinfo{volume}{20}}, \bibinfo{pages}{665} (\bibinfo{year}{1968}), \urlprefix\url{https://link.aps.org/doi/10.1103/PhysRevLett.20.665}.

\bibitem[{\citenamefont{Sampathkumaran and Das}(1995{\natexlab{a}})}]{PhysRevB.51.8631}
\bibinfo{author}{\bibfnamefont{E.~V.} \bibnamefont{Sampathkumaran}} \bibnamefont{and} \bibinfo{author}{\bibfnamefont{I.}~\bibnamefont{Das}}, \bibinfo{journal}{Phys. Rev. B} \textbf{\bibinfo{volume}{51}}, \bibinfo{pages}{8631} (\bibinfo{year}{1995}{\natexlab{a}}), \urlprefix\url{https://link.aps.org/doi/10.1103/PhysRevB.51.8631}.

\bibitem[{\citenamefont{Sampathkumaran and Das}(1995{\natexlab{b}})}]{PhysRevB.51.8178}
\bibinfo{author}{\bibfnamefont{E.~V.} \bibnamefont{Sampathkumaran}} \bibnamefont{and} \bibinfo{author}{\bibfnamefont{I.}~\bibnamefont{Das}}, \bibinfo{journal}{Phys. Rev. B} \textbf{\bibinfo{volume}{51}}, \bibinfo{pages}{8178} (\bibinfo{year}{1995}{\natexlab{b}}), \urlprefix\url{https://link.aps.org/doi/10.1103/PhysRevB.51.8178}.

\bibitem[{\citenamefont{Mallik et~al.}(1997)\citenamefont{Mallik, Paulose, Sampathkumaran, Patil, and Nagarajan}}]{PhysRevB.55.8369}
\bibinfo{author}{\bibfnamefont{R.}~\bibnamefont{Mallik}}, \bibinfo{author}{\bibfnamefont{P.~L.} \bibnamefont{Paulose}}, \bibinfo{author}{\bibfnamefont{E.~V.} \bibnamefont{Sampathkumaran}}, \bibinfo{author}{\bibfnamefont{S.}~\bibnamefont{Patil}}, \bibnamefont{and} \bibinfo{author}{\bibfnamefont{V.}~\bibnamefont{Nagarajan}}, \bibinfo{journal}{Phys. Rev. B} \textbf{\bibinfo{volume}{55}}, \bibinfo{pages}{8369} (\bibinfo{year}{1997}), \urlprefix\url{https://link.aps.org/doi/10.1103/PhysRevB.55.8369}.

\bibitem[{\citenamefont{Mallik and Sampathkumaran}(1998)}]{PhysRevB.58.9178}
\bibinfo{author}{\bibfnamefont{R.}~\bibnamefont{Mallik}} \bibnamefont{and} \bibinfo{author}{\bibfnamefont{E.~V.} \bibnamefont{Sampathkumaran}}, \bibinfo{journal}{Phys. Rev. B} \textbf{\bibinfo{volume}{58}}, \bibinfo{pages}{9178} (\bibinfo{year}{1998}), \urlprefix\url{https://link.aps.org/doi/10.1103/PhysRevB.58.9178}.

\bibitem[{\citenamefont{Chandragiri et~al.}(2016)\citenamefont{Chandragiri, Iyer, and Sampathkumaran}}]{Chandragiri_2016}
\bibinfo{author}{\bibfnamefont{V.}~\bibnamefont{Chandragiri}}, \bibinfo{author}{\bibfnamefont{K.~K.} \bibnamefont{Iyer}}, \bibnamefont{and} \bibinfo{author}{\bibfnamefont{E.~V.} \bibnamefont{Sampathkumaran}}, \bibinfo{journal}{Journal of Physics: Condensed Matter} \textbf{\bibinfo{volume}{28}}, \bibinfo{pages}{286002} (\bibinfo{year}{2016}), \urlprefix\url{https://dx.doi.org/10.1088/0953-8984/28/28/286002}.

\bibitem[{\citenamefont{Pecharsky and Gschneidner}(1997)}]{PhysRevLett.78.4494}
\bibinfo{author}{\bibfnamefont{V.~K.} \bibnamefont{Pecharsky}} \bibnamefont{and} \bibinfo{author}{\bibfnamefont{K.~A.} \bibnamefont{Gschneidner}, \bibfnamefont{Jr.}}, \bibinfo{journal}{Phys. Rev. Lett.} \textbf{\bibinfo{volume}{78}}, \bibinfo{pages}{4494} (\bibinfo{year}{1997}), \urlprefix\url{https://link.aps.org/doi/10.1103/PhysRevLett.78.4494}.

\bibitem[{\citenamefont{Tishin and Spichkin}(2016)}]{tishin2016magnetocaloric}
\bibinfo{author}{\bibfnamefont{A.~M.} \bibnamefont{Tishin}} \bibnamefont{and} \bibinfo{author}{\bibfnamefont{Y.~I.} \bibnamefont{Spichkin}}, \emph{\bibinfo{title}{The magnetocaloric effect and its applications}} (\bibinfo{publisher}{CRC Press}, \bibinfo{year}{2016}).

\bibitem[{\citenamefont{Pecharsky and {Gschneidner Jr}}(1999)}]{PECHARSKY199944}
\bibinfo{author}{\bibfnamefont{V.~K.} \bibnamefont{Pecharsky}} \bibnamefont{and} \bibinfo{author}{\bibfnamefont{K.~A.} \bibnamefont{{Gschneidner Jr}}}, \bibinfo{journal}{Journal of Magnetism and Magnetic Materials} \textbf{\bibinfo{volume}{200}}, \bibinfo{pages}{44} (\bibinfo{year}{1999}), ISSN \bibinfo{issn}{0304-8853}, \urlprefix\url{https://www.sciencedirect.com/science/article/pii/S0304885399003972}.

\bibitem[{\citenamefont{Li and Yan}(2020)}]{LI2020153810}
\bibinfo{author}{\bibfnamefont{L.}~\bibnamefont{Li}} \bibnamefont{and} \bibinfo{author}{\bibfnamefont{M.}~\bibnamefont{Yan}}, \bibinfo{journal}{Journal of Alloys and Compounds} \textbf{\bibinfo{volume}{823}}, \bibinfo{pages}{153810} (\bibinfo{year}{2020}), ISSN \bibinfo{issn}{0925-8388}, \urlprefix\url{https://www.sciencedirect.com/science/article/pii/S0925838820301730}.

\bibitem[{\citenamefont{Sampathkumaran et~al.}(2003)\citenamefont{Sampathkumaran, Sengupta, Rayaprol, Iyer, Doert, and Jemetio}}]{PhysRevLett.91.036603}
\bibinfo{author}{\bibfnamefont{E.~V.} \bibnamefont{Sampathkumaran}}, \bibinfo{author}{\bibfnamefont{K.}~\bibnamefont{Sengupta}}, \bibinfo{author}{\bibfnamefont{S.}~\bibnamefont{Rayaprol}}, \bibinfo{author}{\bibfnamefont{K.~K.} \bibnamefont{Iyer}}, \bibinfo{author}{\bibfnamefont{T.}~\bibnamefont{Doert}}, \bibnamefont{and} \bibinfo{author}{\bibfnamefont{J.~P.~F.} \bibnamefont{Jemetio}}, \bibinfo{journal}{Phys. Rev. Lett.} \textbf{\bibinfo{volume}{91}}, \bibinfo{pages}{036603} (\bibinfo{year}{2003}), \urlprefix\url{https://link.aps.org/doi/10.1103/PhysRevLett.91.036603}.

\bibitem[{\citenamefont{Kumar et~al.}(2019{\natexlab{a}})\citenamefont{Kumar, Sharma, Iyer, and Sampathkumaran}}]{KUMAR2019165515}
\bibinfo{author}{\bibfnamefont{R.}~\bibnamefont{Kumar}}, \bibinfo{author}{\bibfnamefont{J.}~\bibnamefont{Sharma}}, \bibinfo{author}{\bibfnamefont{K.~K.} \bibnamefont{Iyer}}, \bibnamefont{and} \bibinfo{author}{\bibfnamefont{E.}~\bibnamefont{Sampathkumaran}}, \bibinfo{journal}{Journal of Magnetism and Magnetic Materials} \textbf{\bibinfo{volume}{490}}, \bibinfo{pages}{165515} (\bibinfo{year}{2019}{\natexlab{a}}), ISSN \bibinfo{issn}{0304-8853}, \urlprefix\url{https://www.sciencedirect.com/science/article/pii/S0304885319318748}.

\bibitem[{\citenamefont{Kumar and Sampathkumaran}(2021)}]{KUMAR2021168285}
\bibinfo{author}{\bibfnamefont{R.}~\bibnamefont{Kumar}} \bibnamefont{and} \bibinfo{author}{\bibfnamefont{E.}~\bibnamefont{Sampathkumaran}}, \bibinfo{journal}{Journal of Magnetism and Magnetic Materials} \textbf{\bibinfo{volume}{538}}, \bibinfo{pages}{168285} (\bibinfo{year}{2021}), ISSN \bibinfo{issn}{0304-8853}.

\bibitem[{\citenamefont{Kumar et~al.}(2021)\citenamefont{Kumar, Iyer, Paulose, and Sampathkumaran}}]{PhysRevMaterials.5.054407}
\bibinfo{author}{\bibfnamefont{R.}~\bibnamefont{Kumar}}, \bibinfo{author}{\bibfnamefont{K.~K.} \bibnamefont{Iyer}}, \bibinfo{author}{\bibfnamefont{P.~L.} \bibnamefont{Paulose}}, \bibnamefont{and} \bibinfo{author}{\bibfnamefont{E.~V.} \bibnamefont{Sampathkumaran}}, \bibinfo{journal}{Phys. Rev. Mater.} \textbf{\bibinfo{volume}{5}}, \bibinfo{pages}{054407} (\bibinfo{year}{2021}), \urlprefix\url{https://link.aps.org/doi/10.1103/PhysRevMaterials.5.054407}.

\bibitem[{\citenamefont{Kumar et~al.}(2019{\natexlab{b}})\citenamefont{Kumar, Iyer, Paulose, and Sampathkumaran}}]{10.1063/1.5121293}
\bibinfo{author}{\bibfnamefont{R.}~\bibnamefont{Kumar}}, \bibinfo{author}{\bibfnamefont{K.~K.} \bibnamefont{Iyer}}, \bibinfo{author}{\bibfnamefont{P.~L.} \bibnamefont{Paulose}}, \bibnamefont{and} \bibinfo{author}{\bibfnamefont{E.~V.} \bibnamefont{Sampathkumaran}}, \bibinfo{journal}{Journal of Applied Physics} \textbf{\bibinfo{volume}{126}}, \bibinfo{pages}{123906} (\bibinfo{year}{2019}{\natexlab{b}}), ISSN \bibinfo{issn}{0021-8979}.

\bibitem[{\citenamefont{Kumar et~al.}(2020{\natexlab{b}})\citenamefont{Kumar, Iyer, Paulose, and Sampathkumaran}}]{10.1063/5.0016724}
\bibinfo{author}{\bibfnamefont{R.}~\bibnamefont{Kumar}}, \bibinfo{author}{\bibfnamefont{K.~K.} \bibnamefont{Iyer}}, \bibinfo{author}{\bibfnamefont{P.~L.} \bibnamefont{Paulose}}, \bibnamefont{and} \bibinfo{author}{\bibfnamefont{E.~V.} \bibnamefont{Sampathkumaran}}, \bibinfo{journal}{AIP Conference Proceedings} \textbf{\bibinfo{volume}{2265}}, \bibinfo{pages}{030509} (\bibinfo{year}{2020}{\natexlab{b}}), ISSN \bibinfo{issn}{0094-243X}.

\bibitem[{\citenamefont{Iyer et~al.}(2023)\citenamefont{Iyer, Rayaprol, Kumar, Matteppanavar, Dodamani, Maiti, and Sampathkumaran}}]{magnetochemistry9030085}
\bibinfo{author}{\bibfnamefont{K.~K.} \bibnamefont{Iyer}}, \bibinfo{author}{\bibfnamefont{S.}~\bibnamefont{Rayaprol}}, \bibinfo{author}{\bibfnamefont{R.}~\bibnamefont{Kumar}}, \bibinfo{author}{\bibfnamefont{S.}~\bibnamefont{Matteppanavar}}, \bibinfo{author}{\bibfnamefont{S.}~\bibnamefont{Dodamani}}, \bibinfo{author}{\bibfnamefont{K.}~\bibnamefont{Maiti}}, \bibnamefont{and} \bibinfo{author}{\bibfnamefont{E.~V.} \bibnamefont{Sampathkumaran}}, \bibinfo{journal}{Magnetochemistry} \textbf{\bibinfo{volume}{9}} (\bibinfo{year}{2023}), ISSN \bibinfo{issn}{2312-7481}, \urlprefix\url{https://www.mdpi.com/2312-7481/9/3/85}.

\bibitem[{\citenamefont{Mazumdar et~al.}(1996)\citenamefont{Mazumdar, Nigam, Nagarajan, Godart, Gupta, Padalia, Chandra, and Vijayaraghavan}}]{10.1063/1.115758}
\bibinfo{author}{\bibfnamefont{C.}~\bibnamefont{Mazumdar}}, \bibinfo{author}{\bibfnamefont{A.~K.} \bibnamefont{Nigam}}, \bibinfo{author}{\bibfnamefont{R.}~\bibnamefont{Nagarajan}}, \bibinfo{author}{\bibfnamefont{C.}~\bibnamefont{Godart}}, \bibinfo{author}{\bibfnamefont{L.~C.} \bibnamefont{Gupta}}, \bibinfo{author}{\bibfnamefont{B.~D.} \bibnamefont{Padalia}}, \bibinfo{author}{\bibfnamefont{G.}~\bibnamefont{Chandra}}, \bibnamefont{and} \bibinfo{author}{\bibfnamefont{R.}~\bibnamefont{Vijayaraghavan}}, \bibinfo{journal}{Applied Physics Letters} \textbf{\bibinfo{volume}{68}}, \bibinfo{pages}{3647} (\bibinfo{year}{1996}), ISSN \bibinfo{issn}{0003-6951}.

\bibitem[{\citenamefont{Tokura and Kanazawa}(2021)}]{doi:10.1021/acs.chemrev.0c00297}
\bibinfo{author}{\bibfnamefont{Y.}~\bibnamefont{Tokura}} \bibnamefont{and} \bibinfo{author}{\bibfnamefont{N.}~\bibnamefont{Kanazawa}}, \bibinfo{journal}{Chemical Reviews} \textbf{\bibinfo{volume}{121}}, \bibinfo{pages}{2857} (\bibinfo{year}{2021}), \bibinfo{note}{pMID: 33164494}, \eprint{https://doi.org/10.1021/acs.chemrev.0c00297}, \urlprefix\url{https://doi.org/10.1021/acs.chemrev.0c00297}.

\bibitem[{\citenamefont{Vibhakar et~al.}(2023)\citenamefont{Vibhakar, Khalyavin, Moya, Manuel, Orlandi, Lei, Morosan, and Bombardi}}]{PhysRevB.108.L100404}
\bibinfo{author}{\bibfnamefont{A.~M.} \bibnamefont{Vibhakar}}, \bibinfo{author}{\bibfnamefont{D.~D.} \bibnamefont{Khalyavin}}, \bibinfo{author}{\bibfnamefont{J.~M.} \bibnamefont{Moya}}, \bibinfo{author}{\bibfnamefont{P.}~\bibnamefont{Manuel}}, \bibinfo{author}{\bibfnamefont{F.}~\bibnamefont{Orlandi}}, \bibinfo{author}{\bibfnamefont{S.}~\bibnamefont{Lei}}, \bibinfo{author}{\bibfnamefont{E.}~\bibnamefont{Morosan}}, \bibnamefont{and} \bibinfo{author}{\bibfnamefont{A.}~\bibnamefont{Bombardi}}, \bibinfo{journal}{Phys. Rev. B} \textbf{\bibinfo{volume}{108}}, \bibinfo{pages}{L100404} (\bibinfo{year}{2023}), \urlprefix\url{https://link.aps.org/doi/10.1103/PhysRevB.108.L100404}.

\bibitem[{\citenamefont{Yasui et~al.}(2020)\citenamefont{Yasui, Butler, Khanh, Hayami, Nomoto, Hanaguri, Motome, Arita, Arima, Tokura et~al.}}]{Yasui2020}
\bibinfo{author}{\bibfnamefont{Y.}~\bibnamefont{Yasui}}, \bibinfo{author}{\bibfnamefont{C.~J.} \bibnamefont{Butler}}, \bibinfo{author}{\bibfnamefont{N.~D.} \bibnamefont{Khanh}}, \bibinfo{author}{\bibfnamefont{S.}~\bibnamefont{Hayami}}, \bibinfo{author}{\bibfnamefont{T.}~\bibnamefont{Nomoto}}, \bibinfo{author}{\bibfnamefont{T.}~\bibnamefont{Hanaguri}}, \bibinfo{author}{\bibfnamefont{Y.}~\bibnamefont{Motome}}, \bibinfo{author}{\bibfnamefont{R.}~\bibnamefont{Arita}}, \bibinfo{author}{\bibfnamefont{T.-h.} \bibnamefont{Arima}}, \bibinfo{author}{\bibfnamefont{Y.}~\bibnamefont{Tokura}}, \bibnamefont{et~al.}, \bibinfo{journal}{Nature Communications} \textbf{\bibinfo{volume}{11}}, \bibinfo{pages}{5925} (\bibinfo{year}{2020}), ISSN \bibinfo{issn}{2041-1723}, \urlprefix\url{https://doi.org/10.1038/s41467-020-19751-4}.

\bibitem[{\citenamefont{Hirschberger et~al.}(2019)\citenamefont{Hirschberger, Nakajima, Gao, Peng, Kikkawa, Kurumaji, Kriener, Yamasaki, Sagayama, Nakao et~al.}}]{Hirschberger2019-dw}
\bibinfo{author}{\bibfnamefont{M.}~\bibnamefont{Hirschberger}}, \bibinfo{author}{\bibfnamefont{T.}~\bibnamefont{Nakajima}}, \bibinfo{author}{\bibfnamefont{S.}~\bibnamefont{Gao}}, \bibinfo{author}{\bibfnamefont{L.}~\bibnamefont{Peng}}, \bibinfo{author}{\bibfnamefont{A.}~\bibnamefont{Kikkawa}}, \bibinfo{author}{\bibfnamefont{T.}~\bibnamefont{Kurumaji}}, \bibinfo{author}{\bibfnamefont{M.}~\bibnamefont{Kriener}}, \bibinfo{author}{\bibfnamefont{Y.}~\bibnamefont{Yamasaki}}, \bibinfo{author}{\bibfnamefont{H.}~\bibnamefont{Sagayama}}, \bibinfo{author}{\bibfnamefont{H.}~\bibnamefont{Nakao}}, \bibnamefont{et~al.}, \bibinfo{journal}{Nat. Commun.} \textbf{\bibinfo{volume}{10}}, \bibinfo{pages}{5831} (\bibinfo{year}{2019}).

\bibitem[{\citenamefont{Khanh et~al.}(2020)\citenamefont{Khanh, Nakajima, Yu, Gao, Shibata, Hirschberger, Yamasaki, Sagayama, Nakao, Peng et~al.}}]{Khanh2020-ra}
\bibinfo{author}{\bibfnamefont{N.~D.} \bibnamefont{Khanh}}, \bibinfo{author}{\bibfnamefont{T.}~\bibnamefont{Nakajima}}, \bibinfo{author}{\bibfnamefont{X.}~\bibnamefont{Yu}}, \bibinfo{author}{\bibfnamefont{S.}~\bibnamefont{Gao}}, \bibinfo{author}{\bibfnamefont{K.}~\bibnamefont{Shibata}}, \bibinfo{author}{\bibfnamefont{M.}~\bibnamefont{Hirschberger}}, \bibinfo{author}{\bibfnamefont{Y.}~\bibnamefont{Yamasaki}}, \bibinfo{author}{\bibfnamefont{H.}~\bibnamefont{Sagayama}}, \bibinfo{author}{\bibfnamefont{H.}~\bibnamefont{Nakao}}, \bibinfo{author}{\bibfnamefont{L.}~\bibnamefont{Peng}}, \bibnamefont{et~al.}, \bibinfo{journal}{Nat. Nanotechnol.} \textbf{\bibinfo{volume}{15}}, \bibinfo{pages}{444} (\bibinfo{year}{2020}).

\end{thebibliography}

\end{document}